\documentclass[%
superscriptaddress,
reprint,
showpacs,
 amsmath,amssymb,
 aps,
pra,
xcolor
]{revtex4-1}

\usepackage{color}
\usepackage{graphicx}
\usepackage{dcolumn}
\usepackage{bm}
\usepackage{amssymb}

\begin{document}

\title{
Optical and spin-coherence properties of rubidium atoms trapped  in solid neon
}

\author{Ugne Dargyte}
\affiliation{Department of Physics, University of Nevada, Reno NV 89557, USA}
\author{David M. Lancaster}
\affiliation{Department of Physics, University of Nevada, Reno NV 89557, USA}
\author{Jonathan D. Weinstein}
\email{weinstein@physics.unr.edu}
\homepage{http://www.weinsteinlab.org}
\affiliation{Department of Physics, University of Nevada, Reno NV 89557, USA}


\begin{abstract}
In this work, we measure the properties of ensembles of rubidium atoms trapped in solid neon that are relevant for use as quantum sensors of magnetic fields: the spin coherence of the trapped atoms, and the  ability to optically control and measure their spin state. We use the rubidium atoms as an AC magnetometer --- by employing an appropriate dynamical decoupling sequence --- and demonstrate NMR detection of $^{21}$Ne atoms co-trapped in the neon matrix. 
\end{abstract}

\maketitle

\section{Introduction}

{\color{black}
Alkali atoms trapped in solid helium \cite{Weis1996, PhysRevA.60.3867, moroshkin2006spectroscopy} and in solid parahydrogen \cite{upadhyay2016longitudinal, PhysRevA.100.063419, PhysRevB.100.024106, upadhyay2020ultralong} exhibit excellent properties for quantum sensing of magnetic fields \cite{budker2007optical, RevModPhys.89.035002}. In ensemble measurements, it has been shown that it is possible to optically control and measure the spin state of the trapped atoms \cite{PhysRevA.60.3867, moroshkin2006spectroscopy, upadhyay2016longitudinal, PhysRevA.100.063419}, and it was found that the trapped atoms have long spin coherence times \cite{Weis1996, moroshkin2006spectroscopy, PhysRevB.100.024106, upadhyay2020ultralong}.
}
These are promising properties for achieving single-molecule NMR \cite{taylor2008high}. By co-trapping the ``target'' species to be measured within the matrix at high densities, if one could address a single alkali atom within the matrix, one could use it to perform NMR measurements of a single nearby target molecule \cite{upadhyay2020ultralong}. These ideas have previously been demonstrated beautifully with NV centers in diamond, with both NMR measurements of individual nearby $^{13}$C nuclei \cite{zhao2012sensing, kolkowitz2012sensing, taminiau2012detection} as well as the spatial imaging of dozens of neighboring $^{13}$C nuclei within the diamond \cite{abobeih2019atomic}. 

Unfortunately the advantagous properties of alkali atoms in solid helium and parahydrogen are accompanied by technical problems which are disadvantageous for realizing single-molecule NMR. Because helium does not form a solid at low pressures, samples cannot be grown by standard vapor deposition techniques; this makes it difficult to implant an arbitrary target species at high densities \cite{moroshkin2006spectroscopy}. Alkali atoms in parahydrogen have favorable optical properties for ensemble detection \cite{PhysRevA.100.063419}, but the combination of large optical broadening and the failure to date to observe laser-induced fluorescence (LIF) makes single-atom detection a daunting task \cite{PhysRevA.103.052614}.

In this paper, we investigate the properties of rubidium atoms trapped in solid neon. Unlike helium, neon has a zero-pressure solid phase. Therefore it can be grown by vapor deposition and doped at high densities \cite{bondybey1996new}. Unlike parahydrogen, rubidium trapped in solid neon has demonstrated LIF, and has been measured to emit frequency-shifted light with a high quantum efficiency \cite{PhysRevA.103.052614}.

\section{Sample growth}

\begin{figure}[ht]
    \begin{center}
    \includegraphics[width=\linewidth]{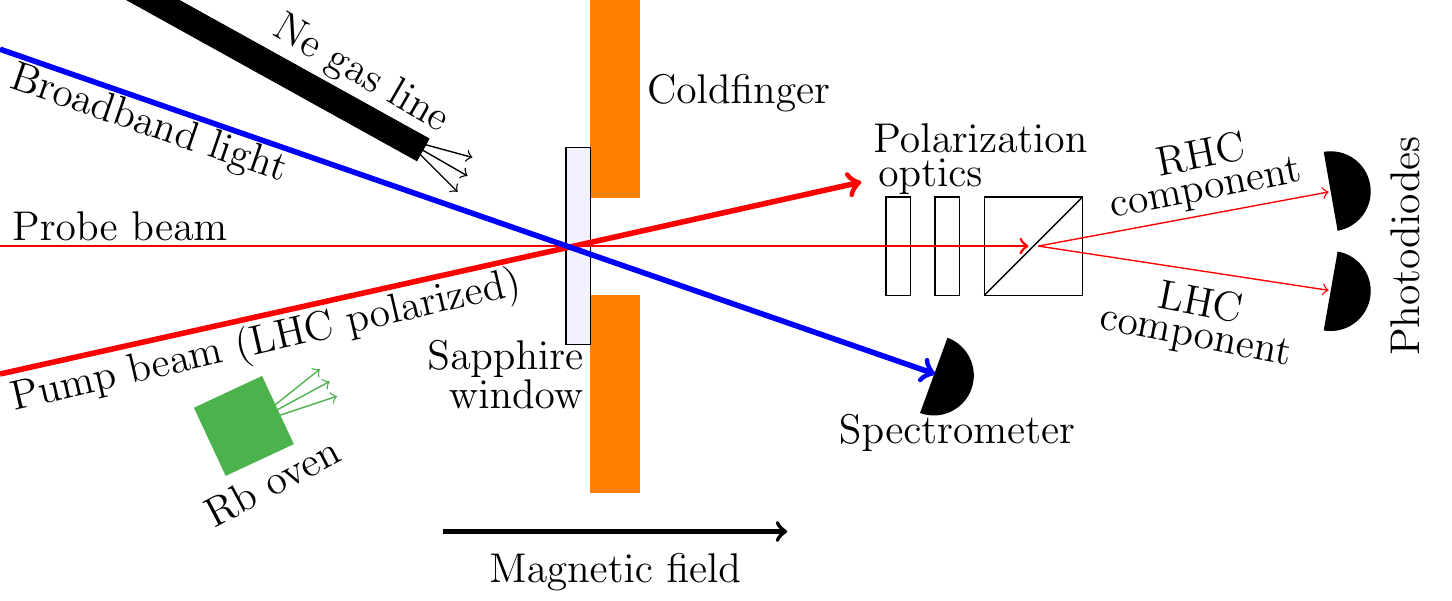}
    \caption{ \color{black} Schematic of the experiment. The static magnetic field is created using permanent magnets outside the vacuum chamber. The rf magnetic field is generated by a wire a few mm above the window surface (to the left in the diagram).
\label{fig:ExperimentSchematic}
    }
    \end{center}
\end{figure}

The samples are grown by simultaneous vacuum deposition of neon and rubidium onto a cryogenically-cooled sapphire window substrate. The apparatus is identical to that previously used for growing rubidium-doped parahydrogen samples \cite{PhysRevA.100.063419}. 
{\color{black}A schematic is shown in Fig. \ref{fig:ExperimentSchematic}. }
The neon gas is introduced through a precooled line that is typically held at $\sim$20 K during deposition, and rubidium is produced from an oven.
Typical neon deposition rates are 0.1~mm per hour, and the dopant density in the solid is determined by the relative fluxes of rubidium and neon. Typical sample thicknesses range from 0.2 to 0.6~mm.
The substrate temperature during growth is measured by a silicon diode thermometer 
directly mounted to the sapphire window; the stated diode accuracy is $\pm 0.1$~K.
The substrate temperature is controlled during growth using a resistive heater. After sample growth the substrate is cooled to its base temperature of 2.9~K and measurements are performed. 

\section{Optical properties}


We measure the optical transmission of our sample via white-light absorption spectroscopy using a halogen lamp and a fiber-coupled grating spectrometer. The transmission of the sample is determined by comparing spectra taken before and after sample deposition. We express the transmission $T$ in terms of the optical depth $OD$ using the definition $T \equiv e^{-OD}$.

\begin{figure}[ht]
    \begin{center}
    \includegraphics[width=\linewidth]{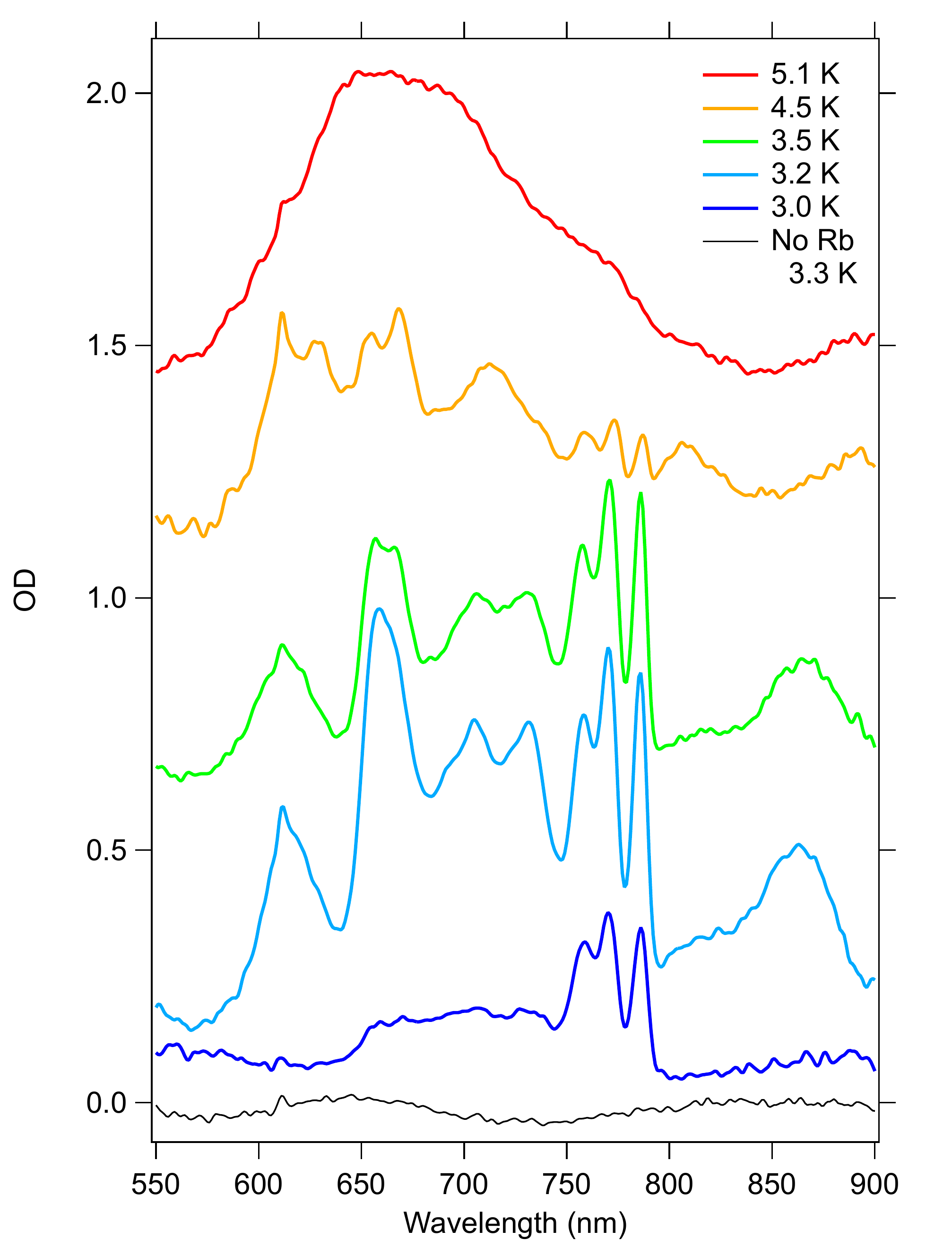}
    \caption{Absorption spectra of samples grown at different substrate temperatures, labeled by the growth temperature. The background scattering from the matrix is subtracted under the assumption that it is a linear interpolation between the OD at 550 and 920~nm. Typical background OD's from scattering by the neon substrate are on the order of 1 in this wavelength range, as seen in the spectrum of Fig. \ref{fig:PolarizationSpectrum}. For clarity, the spectra are vertically offset with higher temperatures plotted above lower temperatures. 
An undoped sample is shown for reference at the bottom.
%
%
%
\label{fig:OpticalSpectra}
    }
    \end{center}
\end{figure}

The optical spectrum of the sample is highly dependent on the substrate temperature during sample growth, as seen in Fig. \ref{fig:OpticalSpectra}. 
Compared to gas-phase rubidium atoms, these spectra exhibit more complicated structure as well as significant line broadening, as is typical for alkali atoms trapped in solid noble-gas matrices \cite{PhysRev.137.A490}. 

%

The spectrum varies most dramatically with substrate temperature, but there are changes due to other growth parameters. Increasing the oven temperature (so as to increase the rubidium flux) resulted in changes to the spectrum qualitatively similar to increasing the substrate temperature.
Varying the neon flow rate by a factor of $\sim 2$ produced little observable change in the atomic absorption spectrum, but increased neon flow rates resulted in significantly larger background scattering from the matrix.





{\color{black}
All optical pumping and probing of the spin state is done with narrow-band light generated by a tunable cw diode laser and a tunable cw titanium-sapphire laser.
}%
We optically pump the spin state of the trapped atoms with a pulse of circularly-polarized light. After pumping we probe the spin state using circular dichroism: we measure the ratio of the transmission of left-hand-circular (LHC) and right-hand-circular (RHC) probe beams \cite{upadhyay2016longitudinal}. We define the polarization signal as the fractional change in this ratio after optical pumping.
Pump beam intensities are on the order of tens of mW/cm$^{2}$, and typical pump pulse durations are on the order of tens of ms. Probe beam intensities are the order of hundreds of $\mu$W/cm$^{2}$. 





\begin{figure}[ht]
    \begin{center}
    \includegraphics[width=\linewidth]{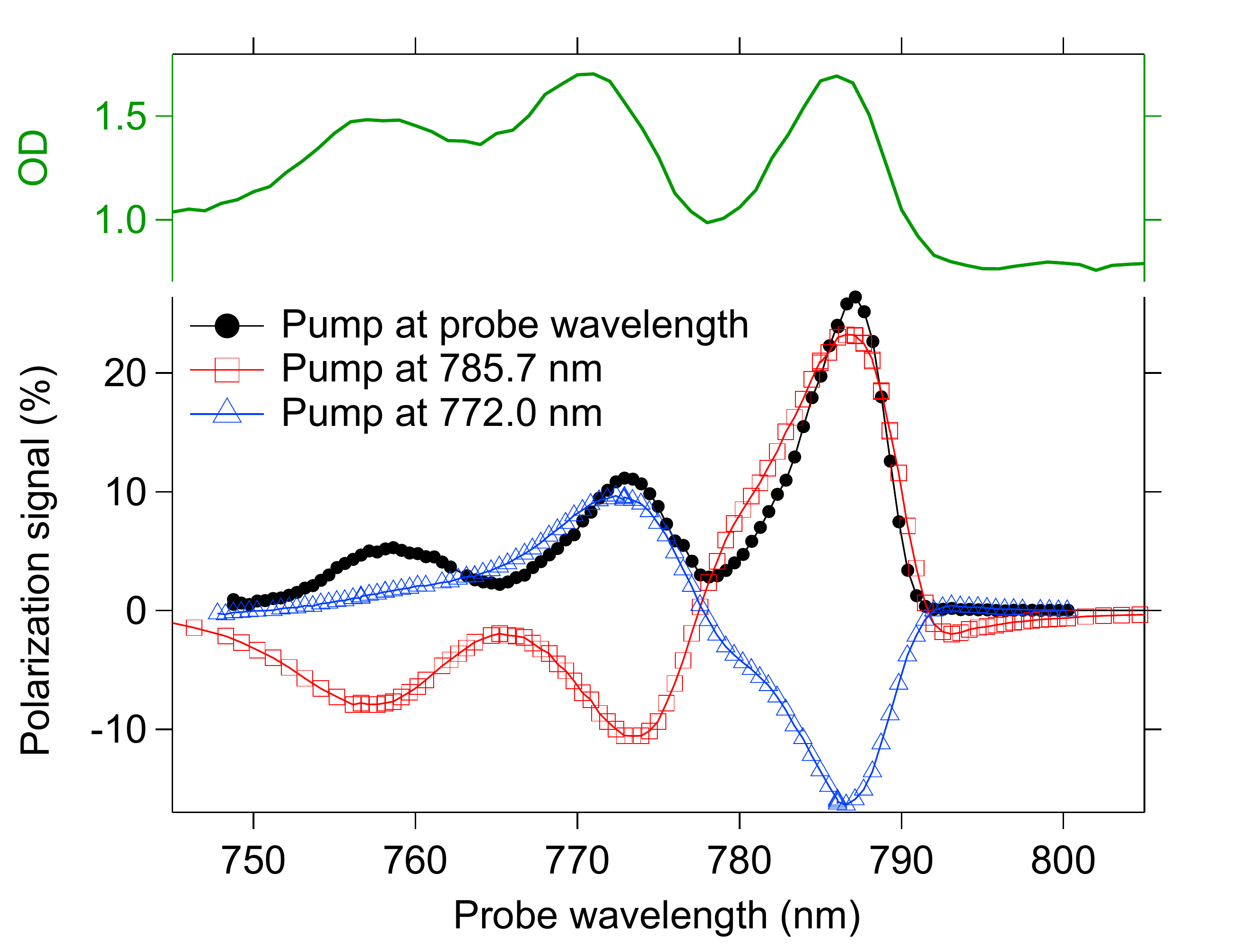}
    \caption{
    The sample polarization signal, as described in the text, as a function of the probe wavelength. Shown is the data for two cases of fixed pump beam frequency (open symbols), as well as the case of degenerate pump and probe beams (filled symbols). 
    At the top, the sample optical depth is plotted for comparison. \label{fig:PolarizationSpectrum}
    }
    \end{center}
\end{figure}

Figure \ref{fig:PolarizationSpectrum} shows the polarization signal obtained for different combinations of pump and probe wavelengths. The sample was grown at a substrate temperature of 3.3~K
and doped at a rubidium density of $1.5 \times 10^{16}$~cm$^{-3}$. The measurements were performed at a magnetic field of 119~G.
Outside of the wavelength range shown in Fig. \ref{fig:PolarizationSpectrum}, no polarization signal was observed when probing the absorption feature further to the infrared. No polarization measurements were made at shorter wavelengths than those shown in Fig. \ref{fig:PolarizationSpectrum} due to laser limitations.
As seen in Fig. \ref{fig:PolarizationSpectrum} the polarization signal exhibits a triplet structure which closely matches that seen in the absorption spectrum, with peaks slightly red-shifted from the absorption spectrum.

We refer to the absorption lines between 750~nm and 800~nm as the ``red triplet''.
A similar triplet was observed by Kupferman and Pipkin for rubidium atoms trapped in solid argon \cite{PhysRev.166.207}. They attributed the triplet structure to the splitting of the $L=1$ excited state due to its ``crystal-field'' interaction with the trapping matrix. Consequently, all three lines of the triplet were believed to originate from a single ``trapping site'' in the matrix.

The data of Fig. \ref{fig:PolarizationSpectrum} confirms this interpretation. Pumping on the rightmost line produces a polarization signal when probing at either of the other two transitions. Consequently the different lines cannot come from different atomic populations.
From the data of Fig. \ref{fig:PolarizationSpectrum} we can similarly conclude that the broadening of each of the three lines is homogenous broadening.

However, as seen in Fig. \ref{fig:OpticalSpectra}, varying the growth conditions causes the height of the red triplet to vary relative to the other absorption features. From this, we conclude that the other lines correspond to a separate population, likely additional trapping sites in the matrix.

Also of note in Fig. \ref{fig:PolarizationSpectrum} is that pumping on the rightmost peak of the red triplet pumps the spin state in the opposite direction as pumping on the middle peak. This indicates that optical pumping is primarily depopulation pumping for at least one of these lines \cite{Happer72OptPumpReview}; this is confirmed by the distribution of $m_F$ states produced, as discussed below in section \ref{sec:t2star}.

For the $\sim 20$ 
samples we have grown and measured, we see a wide variety of polarization signal amplitudes when optically pumping and probing on the rightmost peak of the red triplet. For the samples grown at temperatures from 3.0~K to 4.5~K, we observe that the polarization signal amplitude is linearly proportional to the background-subtracted optical depth of the rightmost peak, to within a standard deviation of $\pm 20\%$. 
(At higher temperatures the background-subtracted optical depth of the red triplet is too small to accurately measure; as expected the polarization signal is also much smaller than for samples grown at lower temperatures.) This observation indicates that the ability to optically pump and probe the spin state of atoms in the red triplet trapping site is independent of the growth temperature, but that low growth temperatures are advantageous because a greater fraction of the implanted rubidium atoms are in the red triplet trapping site.


For the remainder of the paper, we pump and probe on the rightmost peak of the red triplet with degenerate beams generated from a single laser.
Because such measurements will only interact with rubidium atoms in the red triplet trapping site, all rubidium densities quoted in this paper are for that subset of  dopant atoms. Densities are calculated from the sample thickness and the absorption spectrum from $\sim744$~nm to $\sim794$~nm under the assumption that the background is a linear interpolation of the optical depths at the endpoints of that range, {\color{black} and under the assumption that the Einstein $A$ coefficient is unchanged from the free-atom case \cite{NISTAtomicBasic}.} We estimate a typical factor of $\sim 2$ uncertainty in measured densities, as the actual background from the neighboring peaks is unknown.

\section{T$_1$}
\label{sec:t1}

\begin{figure}[ht]
    \begin{center}
        \includegraphics[width=\linewidth]{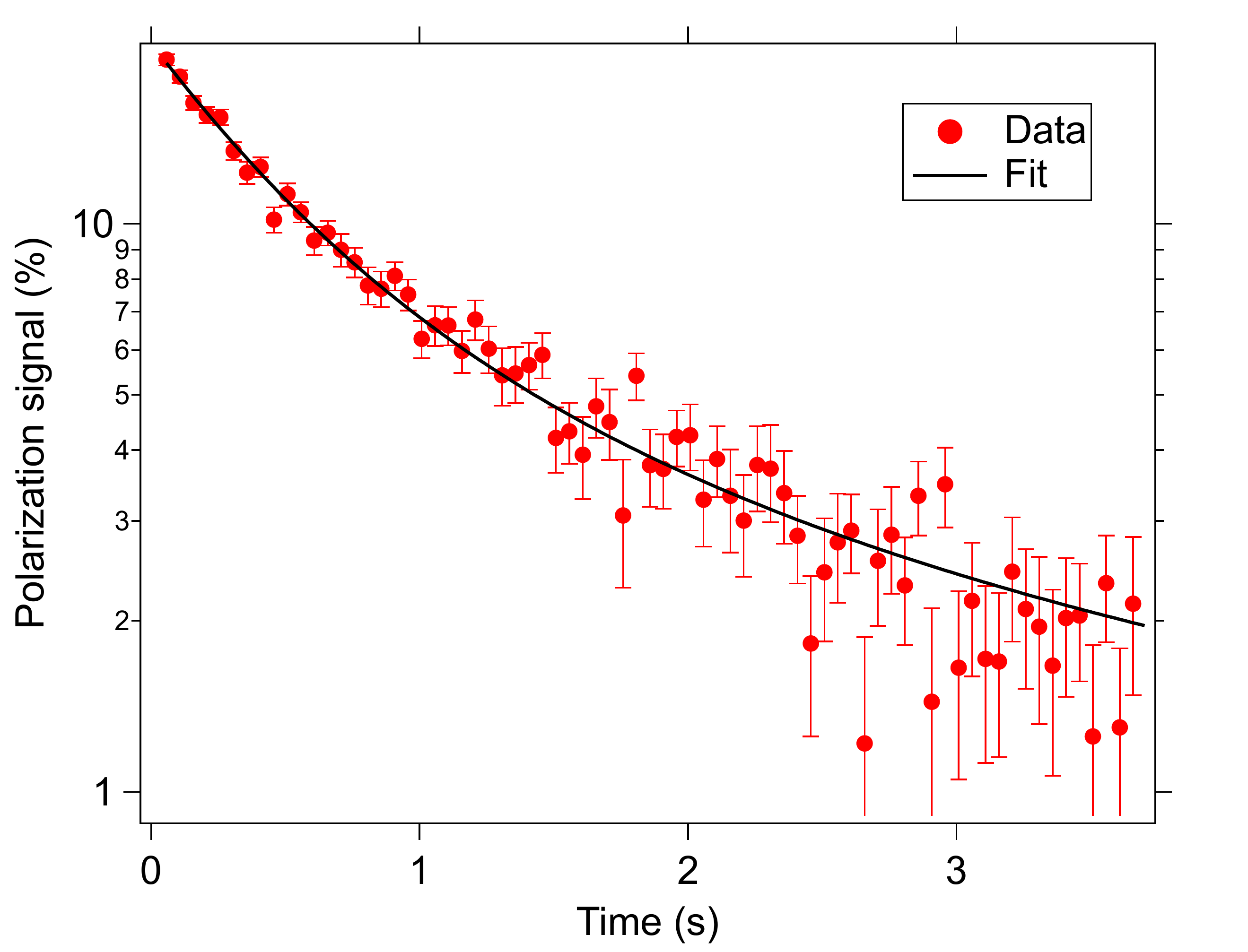}
        \caption{ \color{black} Decay of the polarization signal after the optical pumping beam is turned off at $t=0$. This data is for a sample with a rubidium density of $1.5 \times 10^{16}$~cm$^{-3}$ at a magnetic field of 120~Gauss. Fit as described in the text.
}
        \label{fig:T1}
    \end{center}
\end{figure}

We measure the longitudinal spin relaxation time ($T_1$) by optically pumping to induce a spin polarization and then observing the polarization signal as the spin states return to equilibrium.
{\color{black}
Typical data is shown in Fig. \ref{fig:T1}.

As seen in Fig. \ref{fig:T1}, the decay of the polarization signal is poorly described by a single exponential, which we attribute to inhomogenous $T_1$ times in the sample caused by inhomogenous trapping sites \cite{upadhyay2016longitudinal}. 
We model our atomic ensemble as a uniform distribution of longitudinal relaxation rates from zero to a maximum value, fit the data to determine the maximum relaxation rate, and take the ensemble T$_1$ to be the inverse of the average relaxation rate of the fit \cite{upadhyay2020ultralong}. 
As seen in Fig. \ref{fig:T1}, this simple model fits the data reasonably well.
The data and fit shown in Fig. \ref{fig:T1} yield an ensemble $T_1$ of 0.6~s.
}

Our primary interest in the current work is the coherence time of the trapped rubidium atoms. As $T_1$ is significantly longer than the coherence time, it is not an important limit on the spin coherence, and we did not investigate the details of the physics limiting $T_1$.


\section{Ensemble spin dephasing time (T$_2^*$)}
\label{sec:t2star}

\begin{figure}[ht]
    \begin{center}
        \includegraphics[width=\linewidth]{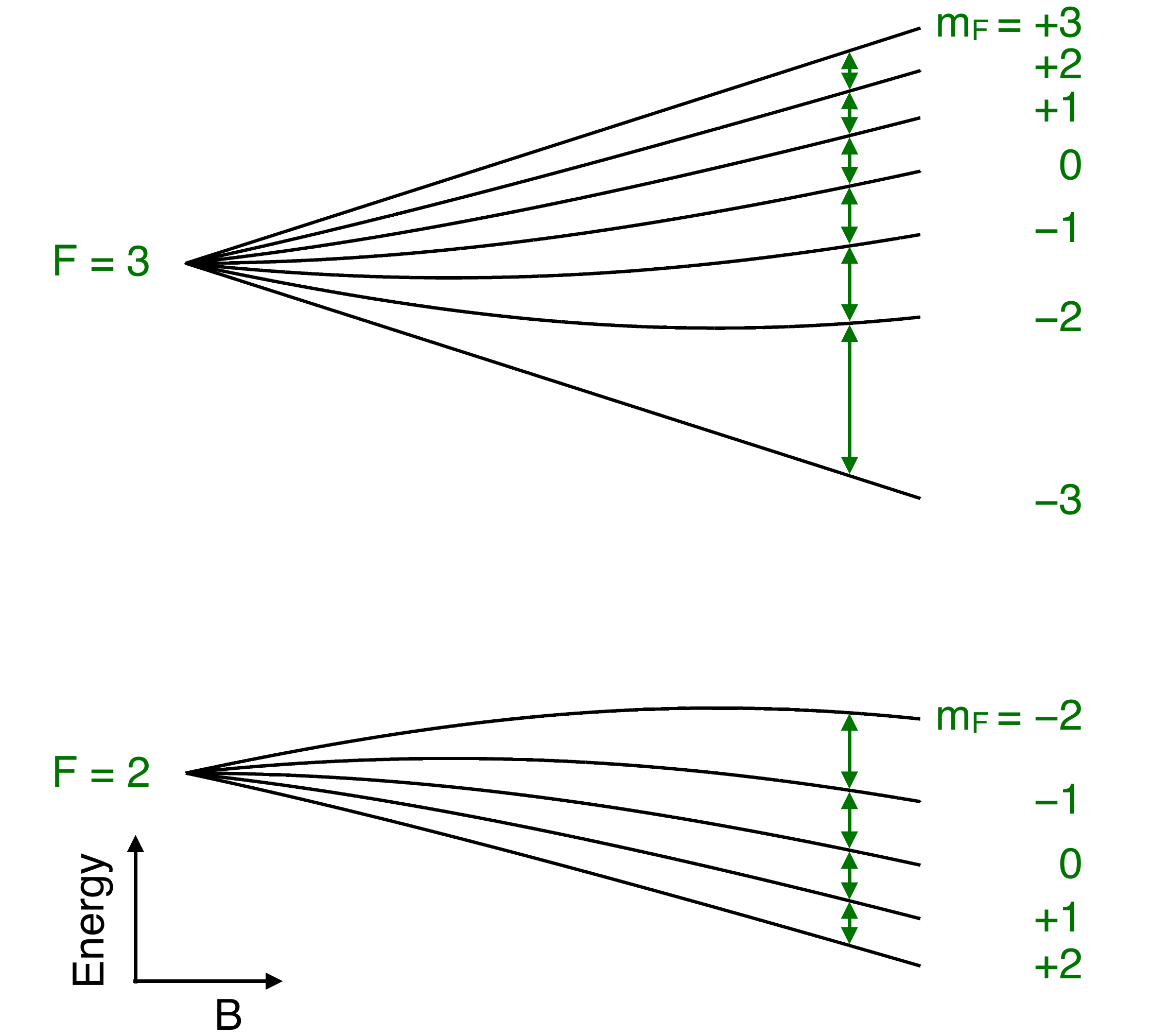}
        \includegraphics[width=\linewidth]{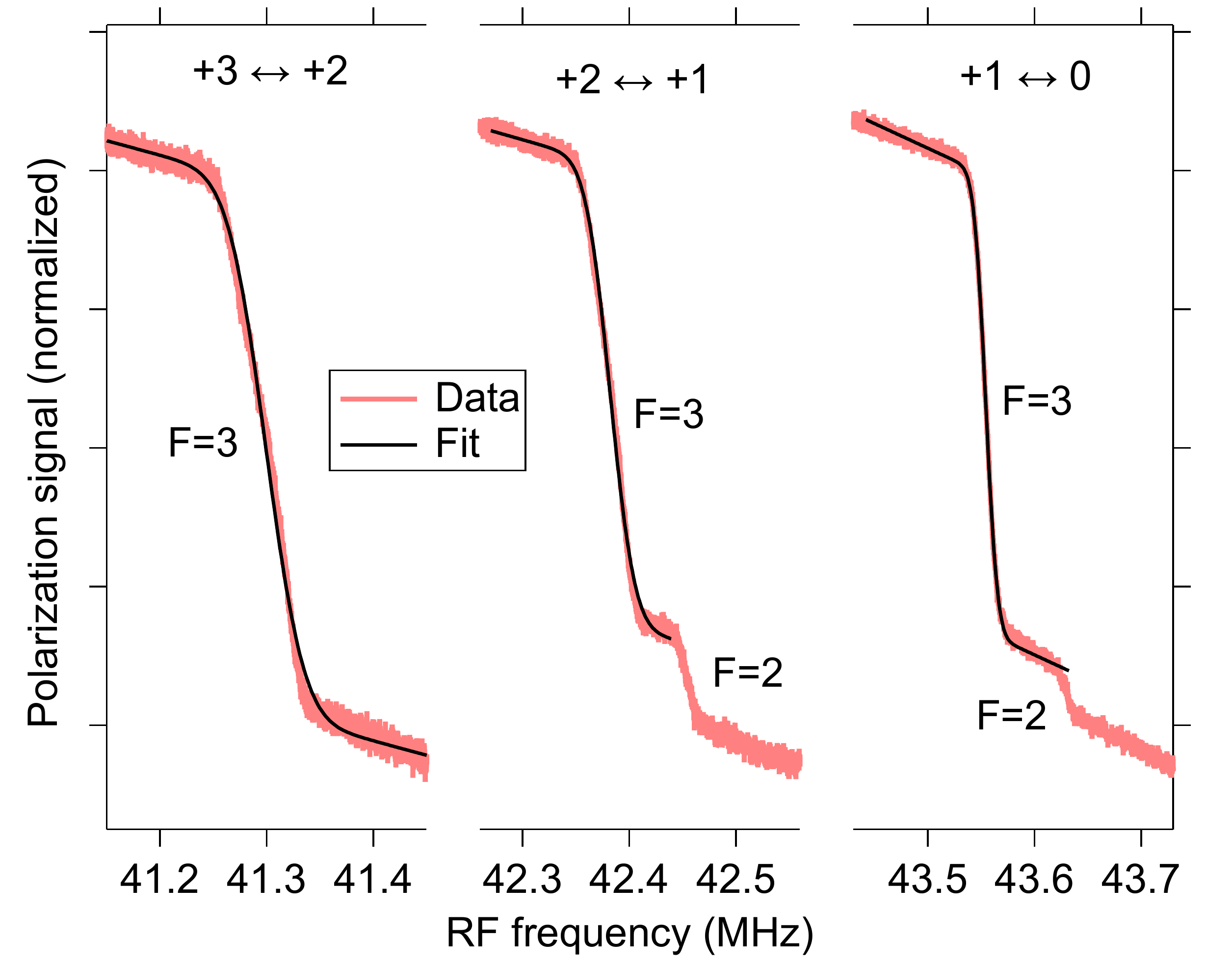}
        \caption{ \color{black} At top, the Zeeman structure of gas-phase $^{85}$Rb in its ground electronic state \cite{arimondo1977experimental}, labeled by the low-field quantum numbers. The relevant Zeeman transitions are shown as arrows (the arrows are placed at a higher magnetic field than used in the experiment to make the nonlinear Zeeman splitting easier to observe). At bottom, an rf spectrum of the Zeeman transitions of $^{85}$Rb, taken at a magnetic field of 95~G, showing a subset of the transitions observed. The fits are used to determine the FWHM linewidths \cite{PhysRevB.100.024106}. 
        The transitions are labeled by their $m_F$ levels and the corresponding $F$ manifold.
}
        \label{fig:T2*raw}
    \end{center}
\end{figure}

We measure T$_2^*$ using rf spectroscopy: after optically pumping, we continuously monitor the polarization signal as we sweep the frequency of an rf magnetic field across the resonant transitions of $^{85}$Rb, as described in Ref. \cite{PhysRevB.100.024106}. 
{\color{black}
We work at a sufficiently large magnetic field such that the nonlinear Zeeman effect allows us to resolve all the single photon transitions between adjacent m$_F$ levels, as seen in Fig. \ref{fig:T2*raw}. The level structure of the electronic ground state of $^{85}$Rb is shown in Fig. \ref{fig:T2*raw} for reference.
}

We determine the magnetic field from the resonant frequencies of the Zeeman transitions under the assumption that the Zeeman splittings are unchanged from the free-atom case \cite{weltner1989magnetic, arimondo1977experimental}. The relative shifts of the different Zeeman transitions suggest that this is the case to within the measured line-broadening.

\begin{figure}[ht]
    \begin{center}
        \includegraphics[width=\linewidth]{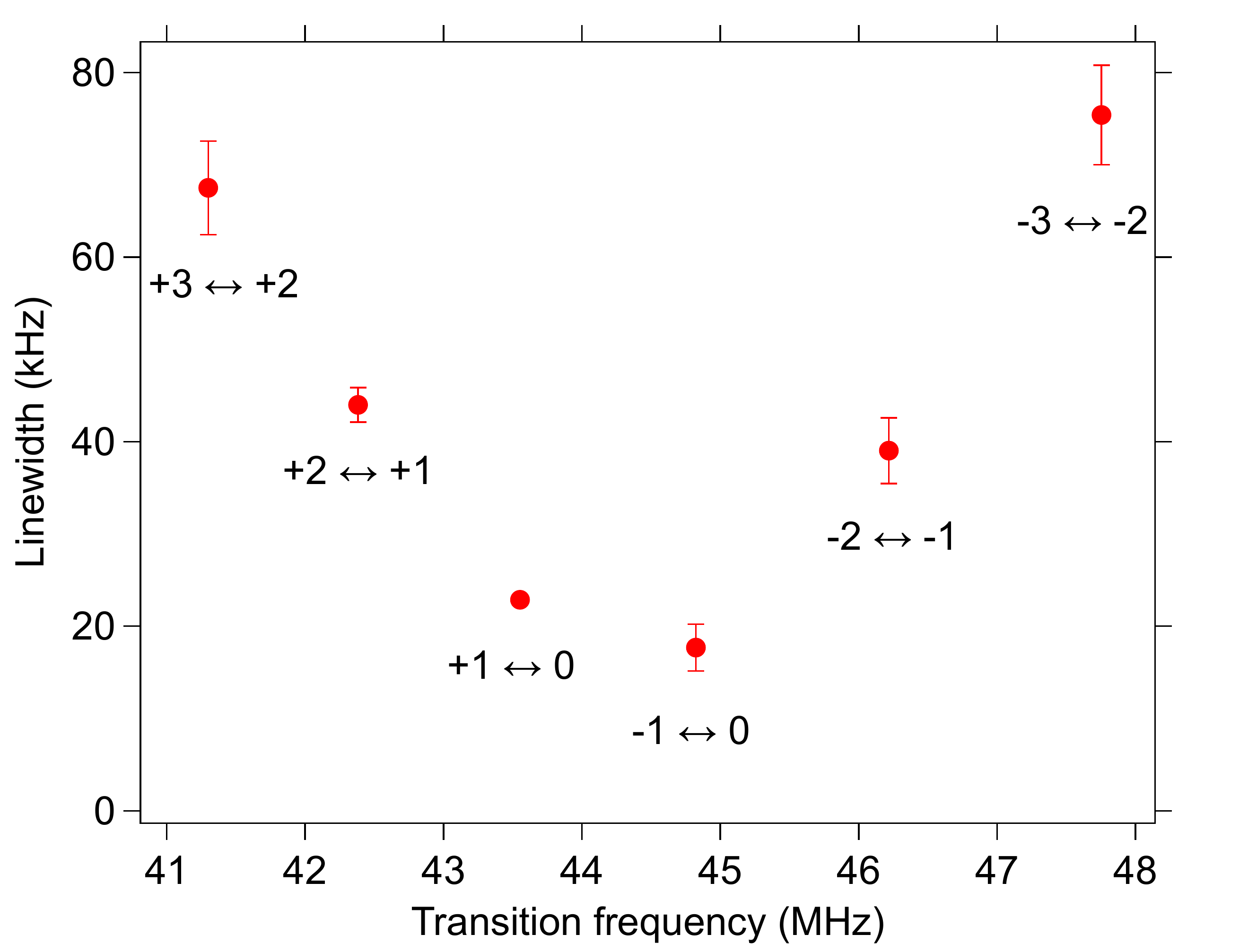}
        \caption{ The full-width-at-half-maximum (FWHM) linewidths of the  transitions between adjacent $m_F$ states in the $F=3$ hyperfine manifold of $^{85}$Rb, labeled accordingly. Taken at a magnetic field of 95 G.
}
        \label{fig:T2*}
    \end{center}
\end{figure}

We can spectrally resolve transitions within both the $F = 2$ and $F = 3$ manifolds of $^{85}$Rb. From the spectra, we find that atoms in the $F = 3$ manifold are optically pumped towards the  $m_F = -3$ level and atoms in the $F = 2$ manifold are pumped towards $m_F = +2$. We conclude that optical pumping of Rb in Ne is primarily depopulation pumping \cite{Happer72OptPumpReview, PhysRevA.100.063419}. This is similar to what was previously reported for Rb atoms trapped in solid helium \cite{PhysRevLett.88.123002}, but contrary to the behavior of Rb in solid parahydrogen, where the optical pumping of spin was found to be predominantly repopulation pumping \cite{PhysRevA.100.063419}. 

Fig. \ref{fig:T2*} shows the measured linewidths for the single photon transitions between $m_F$ states in the $F=3$ manifold.
The corresponding ensemble spin dephasing time $T_2^*$ (in seconds) can be determined from the FWHM linewidths (in Hz) via $T_2^* = (\pi \cdot \mathrm{FWHM}   )^{-1}$. As the $T_2$ values measured with dynamical decoupling (discussed below) are orders-of-magnitude longer than $T_2^*$, we conclude that $T_2^*$ is  predominantly limited by inhomongenous broadening.
By varying the rf power we find that power broadening effects in the data of Fig. \ref{fig:T2*} are at a level $\lesssim 2$~kHz. By comparing to the linewidth of a narrower two-photon transition \cite{PhysRevB.100.024106}, we 
measure the magnetic field inhomogeneity effects to be at a level $\lesssim 7$~kHz.
The linewidths are slighly narrower than what was previously reported for rubidium trapped in parahydrogen, but the ``pattern'' of linewidths is qualitatively similar \cite{PhysRevA.100.063419, PhysRevB.100.024106}.
This suggests that,  as was the case with  parahydrogen, T$_2^*$ is limited primarily by inhomogeneous broadening due to electrostaticlike interactions with the host matrix \cite{PhysRevA.100.063419, PhysRevB.100.024106}.




We found no dependence of T$_2^*$ on the sample growth conditions over the range explored.
We also observed no dependence on the magnetic field: measuring T$_2^*$ at a magnetic field of 119 G gave similar results to those of Fig. \ref{fig:T2*}. 

{\color{black}
It is interesting to compare these measurements to prior $T_2^*$ measurements of the Zeeman transitions of alkali atoms trapped in other noble gas solids.
Cesium atoms trapped in the hcp phase of solid helium have a linewidth on the order of $10^4$~Hz \cite{moroshkin2006spectroscopy}, comparable to our observations of Rb in Ne.
However, cesium trapped in the bcc phase of solid helium exhibit a much narrower linewidth, on the order of $10^1$~Hz, thanks to the symmetry of the bcc phase \cite{Weis1996, moroshkin2006spectroscopy, moroshkin2008atomic}.
Unfortunately, neon and the heavier noble gas solids generally exist in the fcc phase, although there are predictions that the bcc phase can be obtained at high pressures \cite{singh2021solid}. ESR measurements of alkali atoms trapped in heavier noble gases  report broader linewidths than what we find for Rb in Ne, typically by one to two orders of magnitude \cite{goldsborough1964electron, schrimpf1992thermally, vaskonen1999trapping, weltner1989magnetic}. It is unknown whether these differences are due to the matrix species or to other differences in experimental conditions.
}

\section{Ensemble coherence time (T$_2$)}
\label{sec:t2}

To extend the coherence time we use a dynamical decoupling pulse sequence to reduce the sensitivity of the spin state superposition to inhomogenous broadening in the matrix and to environmental noise. As described below, we use the alternating-phase Carr-Purcell (APCP) sequence \cite{PhysRev.94.630, slichter2013principles}, which is of particular interest because it can be used to create an AC magnetometer sensitive to a single frequency (and  harmonics)
\cite{taylor2008high, Farfurnik_2018}.

\begin{figure}[ht]
    \begin{center}
    \includegraphics[width=\linewidth,trim=4 4 4 4,clip]{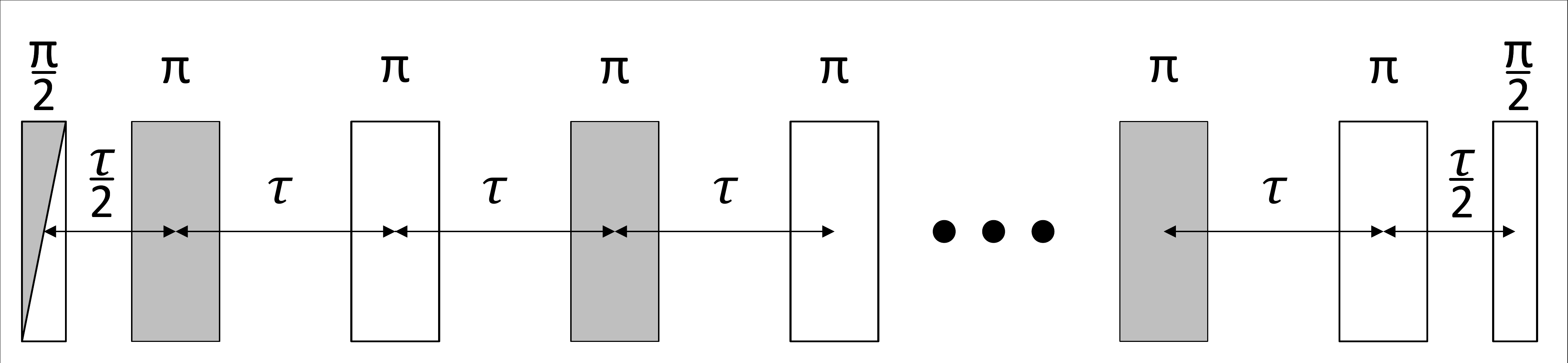}
    \caption{\color{black} Schematic of the APCP sequence, as described in the text. Shading denotes the relative phase of the pulse: the shaded pulses are 180-degrees out of phase with the unshaded pulses. We vary the phase of the first pulse as described in the text.
\label{fig:APCP_Schematic}
    }
    \end{center}
\end{figure}

We measure the coherence lifetime of $^{85}$Rb atoms in a superposition of the  $m_F=-1, 0$ levels of the $F=3$ manifold. The experimental sequence is as follows:
First, we optically pump the atoms as described above, and then apply an rf sweep to maximize the population in the $F=3, m_F=-1$ state.
We then apply a $\pi/2$ pulse on resonance with the $m_F = -1 \leftrightarrow 0$ transition to create the desired superposition state.
{\color{black}
This is followed by the APCP sequence: $(\frac{\tau}{2} - \pi - \tau - \pi - \frac{\tau}{2})^N$, where the $\tau$ terms denote waiting times, $\pi$ denotes a $\pi$-pulse on resonance with the $m_F = -1 \leftrightarrow 0$ transition, and the sequence is repeated $N$ times. The phase of sequential $\pi$ pulses alternates by 180 degrees to correct for rotation errors, as shown in Figure \ref{fig:APCP_Schematic}.
}

All rf pulses in the sequence are generated by a single arbitrary waveform generator for phase accuracy. When taking data, we observe that the transition frequency drifts on a timescale of tens of hours, which we attribute to magnetic field instabilities in the laboratory. We perform rf spectroscopy of the transition every few hours to re-measure the transition frequency and ensure the APCP pulses are on resonance.

\begin{figure}[ht]
    \begin{center}
    \includegraphics[width=\linewidth]{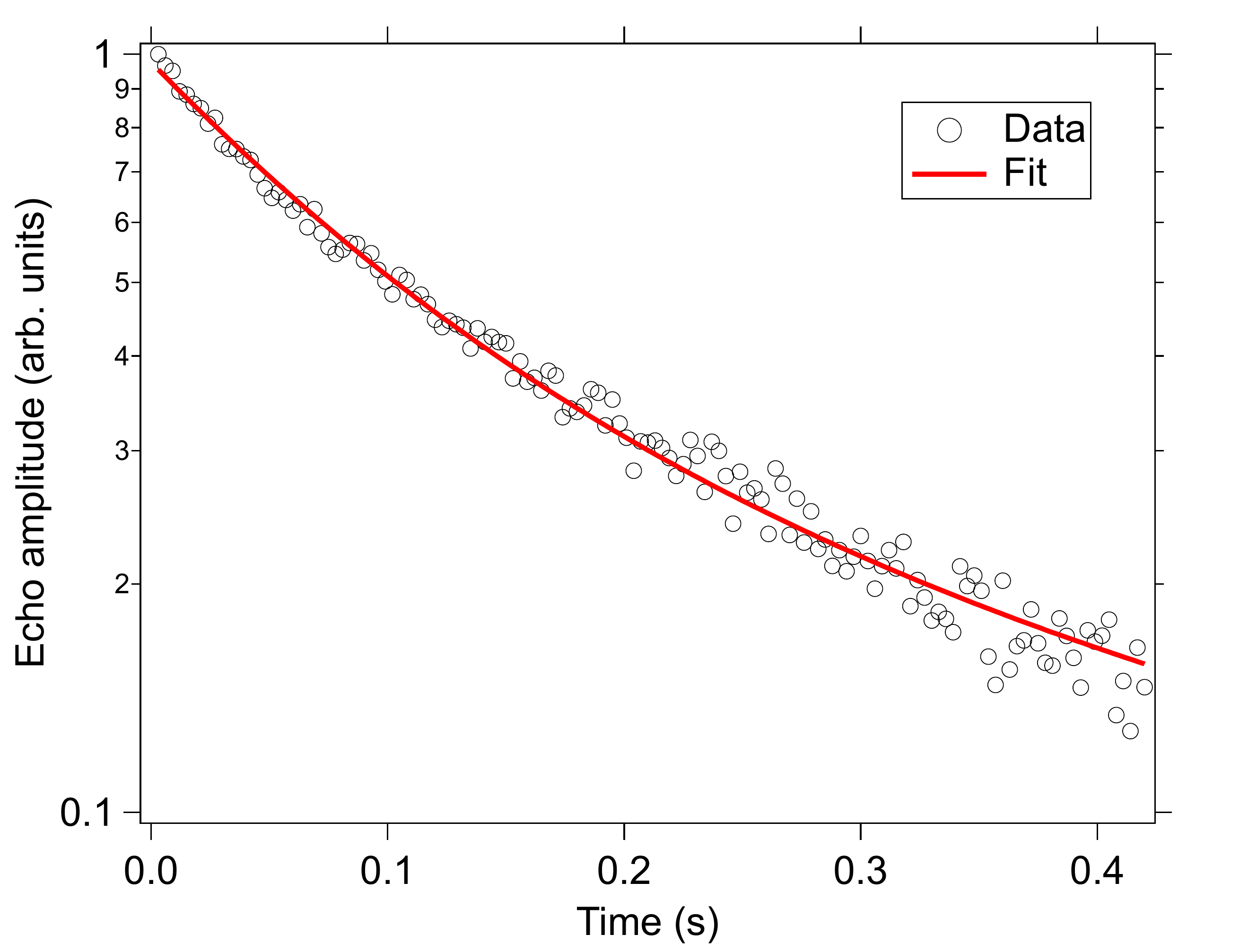}
    \caption{ APCP signal, as described in the text.
    The data was taken at a magnetic field of 119~G, the $\pi$-pulse duration was $2.5~\mu$s, and the $\pi$-pulse repetition rate was 0.3~MHz. The fit, as described in the text, gives an ensemble $T_2$ of 0.14~s. The density of rubidium atoms in the red triplet trapping site was $5 \times 10^{15}$~cm$^{-3}$.
\label{fig:APCP_time}
    }
    \end{center}
\end{figure}

The APCP sequence generates a series of spin echoes, which we observe optically as in Ref. \cite{upadhyay2020ultralong}. To ensure that we are measuring the echoes of the original $\pi/2$ pulse, and not some artifact of imperfect pulses, we repeat the experimental sequence for opposite phases of the initial $\pi/2$ pulse, and subtract the two resulting signals. 
The resulting data is shown in Fig.~\ref{fig:APCP_time}.

We fit the decay of the echo amplitude to extract the ensemble spin coherence lifetime T$_2$. As seen in figure~\ref{fig:APCP_time}, the decay curve fits poorly to an exponential. This is similar to what was observed for prior work measuring the Rb spin coherence in solid parahydrogen \cite{upadhyay2020ultralong}. This is not surprising, as we expect an inhomogenous distribution of trapping sites (and distances to nearest-neighbor magnetic impurities in the matrix) giving rise to an inhomogenous distribution of decay times for the atoms in the ensemble.

{\color{black}
Similar to the analysis of $T_1$ in section \ref{sec:t1}, we model our atomic ensemble as a uniform distribution of decoherence rates from zero to a maximum value, fit the data to determine the maximum decoherence rate, and take the ensemble T$_2$ to be the inverse of the average decoherence rate in the fit \cite{upadhyay2020ultralong}. 
As seen in Fig. \ref{fig:APCP_time}, this simple model fits the data reasonably well, with some systematic deviations at early times (likely due to atoms with decoherence rates larger than the maximum of the model's flat distribution) and late times (likely due to an absence of atoms with near-zero decoherence rates).
}

{\color{black}
The measured coherence times are longest at high $\pi$-pulse repetition rates; Fig. \ref{fig:APCP_time} shows some of the highest-repetition-rate data we were able to obtain, limited by the Rabi frequencies achievable in our current apparatus. We explore the dependence of the coherence time on the pulse repetition rate in detail in Section \ref{sec:APCPspectroscopy}.
}
We note this $T_2$ is comparable to the longest spin-coherence times previously observed for alkali-metal atoms trapped in solid parahydrogen \cite{upadhyay2020ultralong} or helium \cite{Weis1996,  moroshkin2006spectroscopy}.

We also note that achieving long $T_2$ times requires working at low rubidium densities. Repeating the measurements of Fig. \ref{fig:APCP_time} with a sample doped at a higher density ($2 \times 10^{16}$~cm$^{-3}$ Rb atoms in the red triplet trapping site) demonstrated a significantly shorter $T_2$ of 0.01~s.





%

\section{APCP spectroscopy}
\label{sec:APCPspectroscopy}

\begin{figure}[ht]
    \begin{center}
    \includegraphics[width=\linewidth]{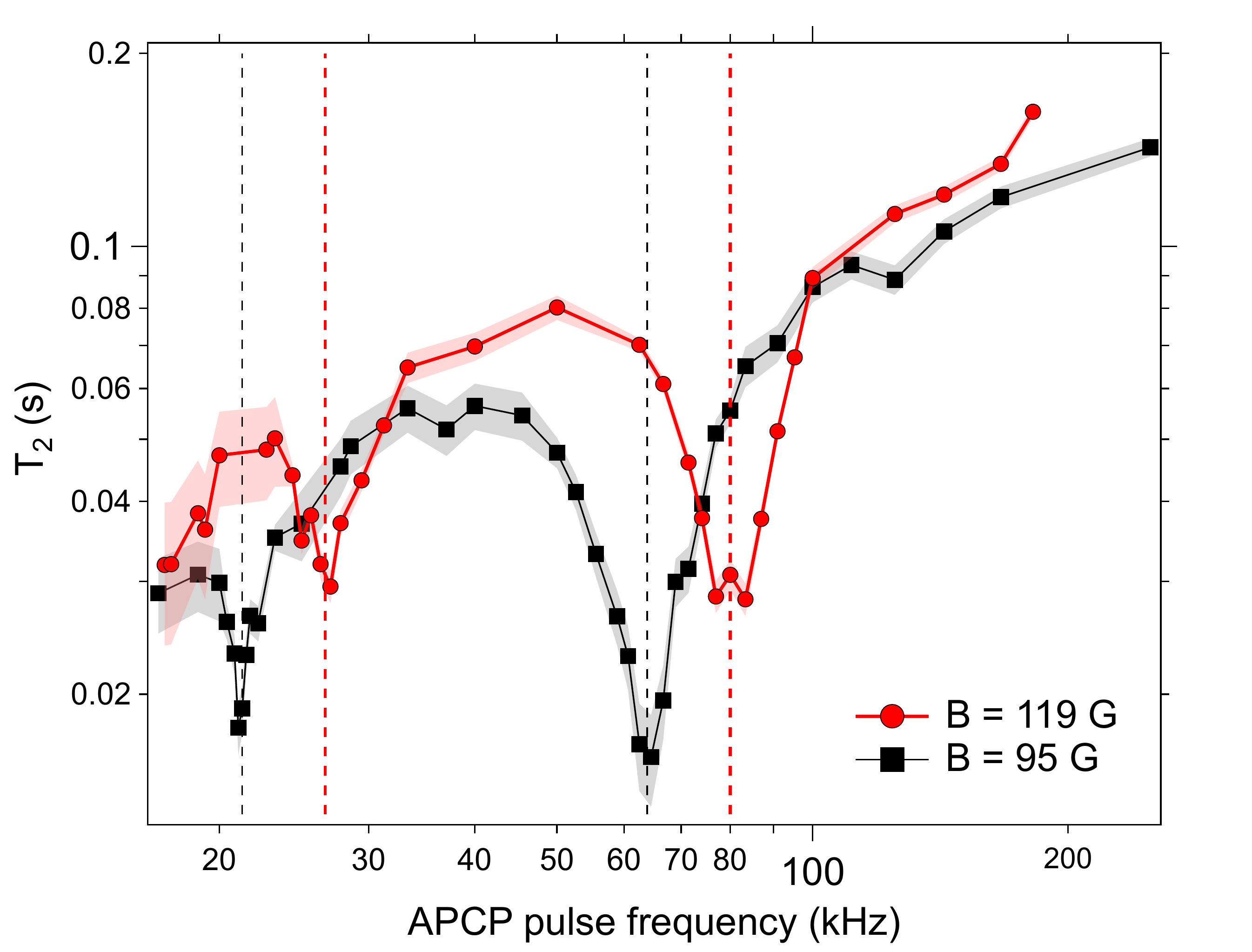}
    \caption{ APCP spectra from the same sample at two different magnetic fields. The dashed vertical lines mark the APCP frequencies at which we would expect to be sensitive to the NMR precession of $^{21}$Ne. 
    The error bands represent the standard deviation of measurements taken over multiple days.
\label{fig:APCP_Spectum}
    }
    \end{center}
\end{figure}

The APCP sequence enables us to use the rubidium spin as a narrow-band AC magnetometer \cite{taylor2008high}. If the $\pi$ pulses of the sequence repeat at a frequency {\color{black}$f_\mathrm{APCP}=1/\tau$ (using the terminology of Fig. \ref{fig:APCP_Schematic})}, the spins are sensitive to magnetic fields at a frequency of $f_\mathrm{APCP}/2$, as well as odd harmonics ($3f_\mathrm{APCP}/2$, $5f_\mathrm{APCP}/2$, \ldots). By repeating the APCP measurement of Fig. \ref{fig:APCP_time} at different values of $f_\mathrm{APCP}$, we obtain a spectrum of the magnetic noise in our sample, in addition to  other sources of decoherence. This data is shown in Fig. \ref{fig:APCP_Spectum} for two different magnetic fields.

From prior work with with NV centers in diamond \cite{childress2006coherent}, rubidium in parahydrogen \cite{upadhyay2020ultralong}, and other systems \cite{PhysRevB.82.121201, PhysRevB.90.241203}, we expect nuclear spins within the solid neon to be a major source of magnetic noise and decoherence. 
$^{21}$Ne is the only naturally-occurring isotope of neon with $I \neq 0$. $^{21}$Ne ($I=3/2$) has a natural abundance of 0.25\% and a gyromagnetic ratio of  3.36 MHz/T \cite{harris2001nmr}.
Due to its quadrupole moment, $^{21}$Ne has a short nuclear spin T$_2$ and T$_2^*$ in the solid: on the order of 1~ms at low temperatures \cite{henry1972pulsed, sirovich1977studies}.

Looking at the data of Fig. \ref{fig:APCP_Spectum}, at both magnetic fields we observe an overall increase in $T_2$ with $f_\mathrm{APCP}$, indicating the presence of significant noise and sources of decoherence at low-frequencies. This is commonly observed for solid-state electron spin systems \cite{NV2013}.
More notable are the ``dips'' in $T_2$. Each spectrum shows two dips corresponding to the NMR precession frequency $f$ of $^{21}$Ne:  the fundamental at  $f_\mathrm{APCP} = 2 f$ and a harmonic at $f_\mathrm{APCP}= 2f/3$.

The dip linewidths are wider than would be expected from prior measurements of the T$_2^*$ of $^{21}$Ne \cite{henry1972pulsed, sirovich1977studies} or from the Rb $T_2$. We attribute this to inhomogenous broadening in the matrix, and expect narrower lines for individual Rb atoms interacting with individual $^{21}$Ne spins \cite{abobeih2018one}.


\section{Conclusions}

When grown under appropriate conditions, rubidium-doped neon samples have favorable properties for the optical pumping and detection of the Rb spin state. By using dynamical decoupling techniques, ultralong spin coherence times can be obtained. This creates an AC magnetometer of sufficient sensitivity to perform NMR spectroscopy of the co-trapped (unpolarized) $^{21}$Ne nuclear spins, even in ensemble measurements.

In future work we hope to use this system to perform NMR measurements of \emph{single} molecules co-trapped within the matrix. This will require reducing the rubidium density so as to optically resolve single rubidium atoms in the matrix. Due to the large broadening of the optical transition (and accompanying reduction in the light-scattering cross section), this will likely require transitioning from  optical readout via absorption to optical readout with laser-induced fluroescence (LIF).
Fortunately, it is known that rubidium atoms trapped in neon emit LIF with high quantum efficiency and are highly resistant to bleaching \cite{PhysRevA.103.052614}. Additionally, prior work has shown that it is possible to optically detect single atoms and single molecules trapped in solid noble gases through LIF \cite{moerner1993optical, moerner1994examining, chambers2019imaging}, although state-sensitive detection has yet to be demonstrated.

\section*{Acknowledgements}
This material is based upon work supported by the National Science Foundation under Grant No. PHY-1912425.
We gratefully acknowledge preliminary measurements and contributions to the experimental apparatus from Andrew N. Kanagin, Pawan Pathak, and Sunil Upadhyay.

U.D. and D.M.L. contributed equally to this work.



\bibliography{RbinNe2021.bib}

\begin{thebibliography}{42}%
\makeatletter
\providecommand \@ifxundefined [1]{%
 \@ifx{#1\undefined}
}%
\providecommand \@ifnum [1]{%
 \ifnum #1\expandafter \@firstoftwo
 \else \expandafter \@secondoftwo
 \fi
}%
\providecommand \@ifx [1]{%
 \ifx #1\expandafter \@firstoftwo
 \else \expandafter \@secondoftwo
 \fi
}%
\providecommand \natexlab [1]{#1}%
\providecommand \enquote  [1]{``#1''}%
\providecommand \bibnamefont  [1]{#1}%
\providecommand \bibfnamefont [1]{#1}%
\providecommand \citenamefont [1]{#1}%
\providecommand \href@noop [0]{\@secondoftwo}%
\providecommand \href [0]{\begingroup \@sanitize@url \@href}%
\providecommand \@href[1]{\@@startlink{#1}\@@href}%
\providecommand \@@href[1]{\endgroup#1\@@endlink}%
\providecommand \@sanitize@url [0]{\catcode `\\12\catcode `\$12\catcode
  `\&12\catcode `\#12\catcode `\^12\catcode `\_12\catcode `\%12\relax}%
\providecommand \@@startlink[1]{}%
\providecommand \@@endlink[0]{}%
\providecommand \url  [0]{\begingroup\@sanitize@url \@url }%
\providecommand \@url [1]{\endgroup\@href {#1}{\urlprefix }}%
\providecommand \urlprefix  [0]{URL }%
\providecommand \Eprint [0]{\href }%
\providecommand \doibase [0]{http://dx.doi.org/}%
\providecommand \selectlanguage [0]{\@gobble}%
\providecommand \bibinfo  [0]{\@secondoftwo}%
\providecommand \bibfield  [0]{\@secondoftwo}%
\providecommand \translation [1]{[#1]}%
\providecommand \BibitemOpen [0]{}%
\providecommand \bibitemStop [0]{}%
\providecommand \bibitemNoStop [0]{.\EOS\space}%
\providecommand \EOS [0]{\spacefactor3000\relax}%
\providecommand \BibitemShut  [1]{\csname bibitem#1\endcsname}%
\let\auto@bib@innerbib\@empty
\bibitem [{\citenamefont {Kanorsky}\ \emph {et~al.}(1996)\citenamefont
  {Kanorsky}, \citenamefont {Lang}, \citenamefont {L\"ucke}, \citenamefont
  {Ross}, \citenamefont {H\"ansch},\ and\ \citenamefont {Weis}}]{Weis1996}%
  \BibitemOpen
  \bibfield  {author} {\bibinfo {author} {\bibfnamefont {S.~I.}\ \bibnamefont
  {Kanorsky}}, \bibinfo {author} {\bibfnamefont {S.}~\bibnamefont {Lang}},
  \bibinfo {author} {\bibfnamefont {S.}~\bibnamefont {L\"ucke}}, \bibinfo
  {author} {\bibfnamefont {S.~B.}\ \bibnamefont {Ross}}, \bibinfo {author}
  {\bibfnamefont {T.~W.}\ \bibnamefont {H\"ansch}}, \ and\ \bibinfo {author}
  {\bibfnamefont {A.}~\bibnamefont {Weis}},\ }\href {\doibase
  10.1103/PhysRevA.54.R1010} {\bibfield  {journal} {\bibinfo  {journal} {Phys.
  Rev. A}\ }\textbf {\bibinfo {volume} {54}},\ \bibinfo {pages} {R1010}
  (\bibinfo {year} {1996})}\BibitemShut {NoStop}%
\bibitem [{\citenamefont {Lang}\ \emph {et~al.}(1999)\citenamefont {Lang},
  \citenamefont {Kanorsky}, \citenamefont {Eichler}, \citenamefont
  {M\"uller-Siebert}, \citenamefont {H\"ansch},\ and\ \citenamefont
  {Weis}}]{PhysRevA.60.3867}%
  \BibitemOpen
  \bibfield  {author} {\bibinfo {author} {\bibfnamefont {S.}~\bibnamefont
  {Lang}}, \bibinfo {author} {\bibfnamefont {S.}~\bibnamefont {Kanorsky}},
  \bibinfo {author} {\bibfnamefont {T.}~\bibnamefont {Eichler}}, \bibinfo
  {author} {\bibfnamefont {R.}~\bibnamefont {M\"uller-Siebert}}, \bibinfo
  {author} {\bibfnamefont {T.~W.}\ \bibnamefont {H\"ansch}}, \ and\ \bibinfo
  {author} {\bibfnamefont {A.}~\bibnamefont {Weis}},\ }\href {\doibase
  10.1103/PhysRevA.60.3867} {\bibfield  {journal} {\bibinfo  {journal} {Phys.
  Rev. A}\ }\textbf {\bibinfo {volume} {60}},\ \bibinfo {pages} {3867}
  (\bibinfo {year} {1999})}\BibitemShut {NoStop}%
\bibitem [{\citenamefont {Moroshkin}\ \emph {et~al.}(2006)\citenamefont
  {Moroshkin}, \citenamefont {Hofer}, \citenamefont {Ulzega},\ and\
  \citenamefont {Weis}}]{moroshkin2006spectroscopy}%
  \BibitemOpen
  \bibfield  {author} {\bibinfo {author} {\bibfnamefont {P.}~\bibnamefont
  {Moroshkin}}, \bibinfo {author} {\bibfnamefont {A.}~\bibnamefont {Hofer}},
  \bibinfo {author} {\bibfnamefont {S.}~\bibnamefont {Ulzega}}, \ and\ \bibinfo
  {author} {\bibfnamefont {A.}~\bibnamefont {Weis}},\ }\href@noop {} {\bibfield
   {journal} {\bibinfo  {journal} {Low Temperature Physics}\ }\textbf {\bibinfo
  {volume} {32}},\ \bibinfo {pages} {981} (\bibinfo {year} {2006})}\BibitemShut
  {NoStop}%
\bibitem [{\citenamefont {Upadhyay}\ \emph {et~al.}(2016)\citenamefont
  {Upadhyay}, \citenamefont {Kanagin}, \citenamefont {Hartzell}, \citenamefont
  {Christy}, \citenamefont {Arnott}, \citenamefont {Momose}, \citenamefont
  {Patterson},\ and\ \citenamefont {Weinstein}}]{upadhyay2016longitudinal}%
  \BibitemOpen
  \bibfield  {author} {\bibinfo {author} {\bibfnamefont {S.}~\bibnamefont
  {Upadhyay}}, \bibinfo {author} {\bibfnamefont {A.~N.}\ \bibnamefont
  {Kanagin}}, \bibinfo {author} {\bibfnamefont {C.}~\bibnamefont {Hartzell}},
  \bibinfo {author} {\bibfnamefont {T.}~\bibnamefont {Christy}}, \bibinfo
  {author} {\bibfnamefont {W.~P.}\ \bibnamefont {Arnott}}, \bibinfo {author}
  {\bibfnamefont {T.}~\bibnamefont {Momose}}, \bibinfo {author} {\bibfnamefont
  {D.}~\bibnamefont {Patterson}}, \ and\ \bibinfo {author} {\bibfnamefont
  {J.~D.}\ \bibnamefont {Weinstein}},\ }\href@noop {} {\bibfield  {journal}
  {\bibinfo  {journal} {Physical Review Letters}\ }\textbf {\bibinfo {volume}
  {117}},\ \bibinfo {pages} {175301} (\bibinfo {year} {2016})}\BibitemShut
  {NoStop}%
\bibitem [{\citenamefont {Upadhyay}\ \emph
  {et~al.}(2019{\natexlab{a}})\citenamefont {Upadhyay}, \citenamefont
  {Dargyte}, \citenamefont {Dergachev}, \citenamefont {Prater}, \citenamefont
  {Varganov}, \citenamefont {Tscherbul}, \citenamefont {Patterson},\ and\
  \citenamefont {Weinstein}}]{PhysRevA.100.063419}%
  \BibitemOpen
  \bibfield  {author} {\bibinfo {author} {\bibfnamefont {S.}~\bibnamefont
  {Upadhyay}}, \bibinfo {author} {\bibfnamefont {U.}~\bibnamefont {Dargyte}},
  \bibinfo {author} {\bibfnamefont {V.~D.}\ \bibnamefont {Dergachev}}, \bibinfo
  {author} {\bibfnamefont {R.~P.}\ \bibnamefont {Prater}}, \bibinfo {author}
  {\bibfnamefont {S.~A.}\ \bibnamefont {Varganov}}, \bibinfo {author}
  {\bibfnamefont {T.~V.}\ \bibnamefont {Tscherbul}}, \bibinfo {author}
  {\bibfnamefont {D.}~\bibnamefont {Patterson}}, \ and\ \bibinfo {author}
  {\bibfnamefont {J.~D.}\ \bibnamefont {Weinstein}},\ }\href {\doibase
  10.1103/PhysRevA.100.063419} {\bibfield  {journal} {\bibinfo  {journal}
  {Phys. Rev. A}\ }\textbf {\bibinfo {volume} {100}},\ \bibinfo {pages}
  {063419} (\bibinfo {year} {2019}{\natexlab{a}})}\BibitemShut {NoStop}%
\bibitem [{\citenamefont {Upadhyay}\ \emph
  {et~al.}(2019{\natexlab{b}})\citenamefont {Upadhyay}, \citenamefont
  {Dargyte}, \citenamefont {Prater}, \citenamefont {Dergachev}, \citenamefont
  {Varganov}, \citenamefont {Tscherbul}, \citenamefont {Patterson},\ and\
  \citenamefont {Weinstein}}]{PhysRevB.100.024106}%
  \BibitemOpen
  \bibfield  {author} {\bibinfo {author} {\bibfnamefont {S.}~\bibnamefont
  {Upadhyay}}, \bibinfo {author} {\bibfnamefont {U.}~\bibnamefont {Dargyte}},
  \bibinfo {author} {\bibfnamefont {R.~P.}\ \bibnamefont {Prater}}, \bibinfo
  {author} {\bibfnamefont {V.~D.}\ \bibnamefont {Dergachev}}, \bibinfo {author}
  {\bibfnamefont {S.~A.}\ \bibnamefont {Varganov}}, \bibinfo {author}
  {\bibfnamefont {T.~V.}\ \bibnamefont {Tscherbul}}, \bibinfo {author}
  {\bibfnamefont {D.}~\bibnamefont {Patterson}}, \ and\ \bibinfo {author}
  {\bibfnamefont {J.~D.}\ \bibnamefont {Weinstein}},\ }\href {\doibase
  10.1103/PhysRevB.100.024106} {\bibfield  {journal} {\bibinfo  {journal}
  {Phys. Rev. B}\ }\textbf {\bibinfo {volume} {100}},\ \bibinfo {pages}
  {024106} (\bibinfo {year} {2019}{\natexlab{b}})}\BibitemShut {NoStop}%
\bibitem [{\citenamefont {Upadhyay}\ \emph {et~al.}(2020)\citenamefont
  {Upadhyay}, \citenamefont {Dargyte}, \citenamefont {Patterson},\ and\
  \citenamefont {Weinstein}}]{upadhyay2020ultralong}%
  \BibitemOpen
  \bibfield  {author} {\bibinfo {author} {\bibfnamefont {S.}~\bibnamefont
  {Upadhyay}}, \bibinfo {author} {\bibfnamefont {U.}~\bibnamefont {Dargyte}},
  \bibinfo {author} {\bibfnamefont {D.}~\bibnamefont {Patterson}}, \ and\
  \bibinfo {author} {\bibfnamefont {J.~D.}\ \bibnamefont {Weinstein}},\
  }\href@noop {} {\bibfield  {journal} {\bibinfo  {journal} {Physical Review
  Letters}\ }\textbf {\bibinfo {volume} {125}},\ \bibinfo {pages} {043601}
  (\bibinfo {year} {2020})}\BibitemShut {NoStop}%
\bibitem [{\citenamefont {Budker}\ and\ \citenamefont
  {Romalis}(2007)}]{budker2007optical}%
  \BibitemOpen
  \bibfield  {author} {\bibinfo {author} {\bibfnamefont {D.}~\bibnamefont
  {Budker}}\ and\ \bibinfo {author} {\bibfnamefont {M.}~\bibnamefont
  {Romalis}},\ }\href@noop {} {\bibfield  {journal} {\bibinfo  {journal}
  {Nature physics}\ }\textbf {\bibinfo {volume} {3}},\ \bibinfo {pages} {227}
  (\bibinfo {year} {2007})}\BibitemShut {NoStop}%
\bibitem [{\citenamefont {Degen}\ \emph {et~al.}(2017)\citenamefont {Degen},
  \citenamefont {Reinhard},\ and\ \citenamefont
  {Cappellaro}}]{RevModPhys.89.035002}%
  \BibitemOpen
  \bibfield  {author} {\bibinfo {author} {\bibfnamefont {C.~L.}\ \bibnamefont
  {Degen}}, \bibinfo {author} {\bibfnamefont {F.}~\bibnamefont {Reinhard}}, \
  and\ \bibinfo {author} {\bibfnamefont {P.}~\bibnamefont {Cappellaro}},\
  }\href {\doibase 10.1103/RevModPhys.89.035002} {\bibfield  {journal}
  {\bibinfo  {journal} {Rev. Mod. Phys.}\ }\textbf {\bibinfo {volume} {89}},\
  \bibinfo {pages} {035002} (\bibinfo {year} {2017})}\BibitemShut {NoStop}%
\bibitem [{\citenamefont {Taylor}\ \emph {et~al.}(2008)\citenamefont {Taylor},
  \citenamefont {Cappellaro}, \citenamefont {Childress}, \citenamefont {Jiang},
  \citenamefont {Budker}, \citenamefont {Hemmer}, \citenamefont {Yacoby},
  \citenamefont {Walsworth},\ and\ \citenamefont {Lukin}}]{taylor2008high}%
  \BibitemOpen
  \bibfield  {author} {\bibinfo {author} {\bibfnamefont {J.}~\bibnamefont
  {Taylor}}, \bibinfo {author} {\bibfnamefont {P.}~\bibnamefont {Cappellaro}},
  \bibinfo {author} {\bibfnamefont {L.}~\bibnamefont {Childress}}, \bibinfo
  {author} {\bibfnamefont {L.}~\bibnamefont {Jiang}}, \bibinfo {author}
  {\bibfnamefont {D.}~\bibnamefont {Budker}}, \bibinfo {author} {\bibfnamefont
  {P.}~\bibnamefont {Hemmer}}, \bibinfo {author} {\bibfnamefont
  {A.}~\bibnamefont {Yacoby}}, \bibinfo {author} {\bibfnamefont
  {R.}~\bibnamefont {Walsworth}}, \ and\ \bibinfo {author} {\bibfnamefont
  {M.}~\bibnamefont {Lukin}},\ }\href@noop {} {\bibfield  {journal} {\bibinfo
  {journal} {Nature Physics}\ }\textbf {\bibinfo {volume} {4}},\ \bibinfo
  {pages} {810} (\bibinfo {year} {2008})}\BibitemShut {NoStop}%
\bibitem [{\citenamefont {Zhao}\ \emph {et~al.}(2012)\citenamefont {Zhao},
  \citenamefont {Honert}, \citenamefont {Schmid}, \citenamefont {Klas},
  \citenamefont {Isoya}, \citenamefont {Markham}, \citenamefont {Twitchen},
  \citenamefont {Jelezko}, \citenamefont {Liu}, \citenamefont {Fedder},\ and\
  \citenamefont {Wrachtrup}}]{zhao2012sensing}%
  \BibitemOpen
  \bibfield  {author} {\bibinfo {author} {\bibfnamefont {N.}~\bibnamefont
  {Zhao}}, \bibinfo {author} {\bibfnamefont {J.}~\bibnamefont {Honert}},
  \bibinfo {author} {\bibfnamefont {B.}~\bibnamefont {Schmid}}, \bibinfo
  {author} {\bibfnamefont {M.}~\bibnamefont {Klas}}, \bibinfo {author}
  {\bibfnamefont {J.}~\bibnamefont {Isoya}}, \bibinfo {author} {\bibfnamefont
  {M.}~\bibnamefont {Markham}}, \bibinfo {author} {\bibfnamefont
  {D.}~\bibnamefont {Twitchen}}, \bibinfo {author} {\bibfnamefont
  {F.}~\bibnamefont {Jelezko}}, \bibinfo {author} {\bibfnamefont {R.-B.}\
  \bibnamefont {Liu}}, \bibinfo {author} {\bibfnamefont {H.}~\bibnamefont
  {Fedder}}, \ and\ \bibinfo {author} {\bibfnamefont {J.}~\bibnamefont
  {Wrachtrup}},\ }\href@noop {} {\bibfield  {journal} {\bibinfo  {journal}
  {Nature nanotechnology}\ }\textbf {\bibinfo {volume} {7}},\ \bibinfo {pages}
  {657} (\bibinfo {year} {2012})}\BibitemShut {NoStop}%
\bibitem [{\citenamefont {Kolkowitz}\ \emph {et~al.}(2012)\citenamefont
  {Kolkowitz}, \citenamefont {Unterreithmeier}, \citenamefont {Bennett},\ and\
  \citenamefont {Lukin}}]{kolkowitz2012sensing}%
  \BibitemOpen
  \bibfield  {author} {\bibinfo {author} {\bibfnamefont {S.}~\bibnamefont
  {Kolkowitz}}, \bibinfo {author} {\bibfnamefont {Q.~P.}\ \bibnamefont
  {Unterreithmeier}}, \bibinfo {author} {\bibfnamefont {S.~D.}\ \bibnamefont
  {Bennett}}, \ and\ \bibinfo {author} {\bibfnamefont {M.~D.}\ \bibnamefont
  {Lukin}},\ }\href@noop {} {\bibfield  {journal} {\bibinfo  {journal}
  {Physical review letters}\ }\textbf {\bibinfo {volume} {109}},\ \bibinfo
  {pages} {137601} (\bibinfo {year} {2012})}\BibitemShut {NoStop}%
\bibitem [{\citenamefont {Taminiau}\ \emph {et~al.}(2012)\citenamefont
  {Taminiau}, \citenamefont {Wagenaar}, \citenamefont {Van~der Sar},
  \citenamefont {Jelezko}, \citenamefont {Dobrovitski},\ and\ \citenamefont
  {Hanson}}]{taminiau2012detection}%
  \BibitemOpen
  \bibfield  {author} {\bibinfo {author} {\bibfnamefont {T.}~\bibnamefont
  {Taminiau}}, \bibinfo {author} {\bibfnamefont {J.}~\bibnamefont {Wagenaar}},
  \bibinfo {author} {\bibfnamefont {T.}~\bibnamefont {Van~der Sar}}, \bibinfo
  {author} {\bibfnamefont {F.}~\bibnamefont {Jelezko}}, \bibinfo {author}
  {\bibfnamefont {V.~V.}\ \bibnamefont {Dobrovitski}}, \ and\ \bibinfo {author}
  {\bibfnamefont {R.}~\bibnamefont {Hanson}},\ }\href@noop {} {\bibfield
  {journal} {\bibinfo  {journal} {Physical review letters}\ }\textbf {\bibinfo
  {volume} {109}},\ \bibinfo {pages} {137602} (\bibinfo {year}
  {2012})}\BibitemShut {NoStop}%
\bibitem [{\citenamefont {Abobeih}\ \emph {et~al.}(2019)\citenamefont
  {Abobeih}, \citenamefont {Randall}, \citenamefont {Bradley}, \citenamefont
  {Bartling}, \citenamefont {Bakker}, \citenamefont {Degen}, \citenamefont
  {Markham}, \citenamefont {Twitchen},\ and\ \citenamefont
  {Taminiau}}]{abobeih2019atomic}%
  \BibitemOpen
  \bibfield  {author} {\bibinfo {author} {\bibfnamefont {M.}~\bibnamefont
  {Abobeih}}, \bibinfo {author} {\bibfnamefont {J.}~\bibnamefont {Randall}},
  \bibinfo {author} {\bibfnamefont {C.}~\bibnamefont {Bradley}}, \bibinfo
  {author} {\bibfnamefont {H.}~\bibnamefont {Bartling}}, \bibinfo {author}
  {\bibfnamefont {M.}~\bibnamefont {Bakker}}, \bibinfo {author} {\bibfnamefont
  {M.}~\bibnamefont {Degen}}, \bibinfo {author} {\bibfnamefont
  {M.}~\bibnamefont {Markham}}, \bibinfo {author} {\bibfnamefont
  {D.}~\bibnamefont {Twitchen}}, \ and\ \bibinfo {author} {\bibfnamefont
  {T.}~\bibnamefont {Taminiau}},\ }\href@noop {} {\bibfield  {journal}
  {\bibinfo  {journal} {Nature}\ }\textbf {\bibinfo {volume} {576}},\ \bibinfo
  {pages} {411} (\bibinfo {year} {2019})}\BibitemShut {NoStop}%
\bibitem [{\citenamefont {Lancaster}\ \emph {et~al.}(2021)\citenamefont
  {Lancaster}, \citenamefont {Dargyte}, \citenamefont {Upadhyay},\ and\
  \citenamefont {Weinstein}}]{PhysRevA.103.052614}%
  \BibitemOpen
  \bibfield  {author} {\bibinfo {author} {\bibfnamefont {D.~M.}\ \bibnamefont
  {Lancaster}}, \bibinfo {author} {\bibfnamefont {U.}~\bibnamefont {Dargyte}},
  \bibinfo {author} {\bibfnamefont {S.}~\bibnamefont {Upadhyay}}, \ and\
  \bibinfo {author} {\bibfnamefont {J.~D.}\ \bibnamefont {Weinstein}},\ }\href
  {\doibase 10.1103/PhysRevA.103.052614} {\bibfield  {journal} {\bibinfo
  {journal} {Phys. Rev. A}\ }\textbf {\bibinfo {volume} {103}},\ \bibinfo
  {pages} {052614} (\bibinfo {year} {2021})}\BibitemShut {NoStop}%
\bibitem [{\citenamefont {Bondybey}\ \emph {et~al.}(1996)\citenamefont
  {Bondybey}, \citenamefont {Smith},\ and\ \citenamefont
  {Agreiter}}]{bondybey1996new}%
  \BibitemOpen
  \bibfield  {author} {\bibinfo {author} {\bibfnamefont {V.~E.}\ \bibnamefont
  {Bondybey}}, \bibinfo {author} {\bibfnamefont {A.~M.}\ \bibnamefont {Smith}},
  \ and\ \bibinfo {author} {\bibfnamefont {J.}~\bibnamefont {Agreiter}},\
  }\href@noop {} {\bibfield  {journal} {\bibinfo  {journal} {Chemical reviews}\
  }\textbf {\bibinfo {volume} {96}},\ \bibinfo {pages} {2113} (\bibinfo {year}
  {1996})}\BibitemShut {NoStop}%
\bibitem [{\citenamefont {Weyhmann}\ and\ \citenamefont
  {Pipkin}(1965)}]{PhysRev.137.A490}%
  \BibitemOpen
  \bibfield  {author} {\bibinfo {author} {\bibfnamefont {W.}~\bibnamefont
  {Weyhmann}}\ and\ \bibinfo {author} {\bibfnamefont {F.~M.}\ \bibnamefont
  {Pipkin}},\ }\href {\doibase 10.1103/PhysRev.137.A490} {\bibfield  {journal}
  {\bibinfo  {journal} {Phys. Rev.}\ }\textbf {\bibinfo {volume} {137}},\
  \bibinfo {pages} {A490} (\bibinfo {year} {1965})}\BibitemShut {NoStop}%
\bibitem [{\citenamefont {Kupferman}\ and\ \citenamefont
  {Pipkin}(1968)}]{PhysRev.166.207}%
  \BibitemOpen
  \bibfield  {author} {\bibinfo {author} {\bibfnamefont {S.~L.}\ \bibnamefont
  {Kupferman}}\ and\ \bibinfo {author} {\bibfnamefont {F.~M.}\ \bibnamefont
  {Pipkin}},\ }\href {\doibase 10.1103/PhysRev.166.207} {\bibfield  {journal}
  {\bibinfo  {journal} {Phys. Rev.}\ }\textbf {\bibinfo {volume} {166}},\
  \bibinfo {pages} {207} (\bibinfo {year} {1968})}\BibitemShut {NoStop}%
\bibitem [{\citenamefont {Happer}(1972)}]{Happer72OptPumpReview}%
  \BibitemOpen
  \bibfield  {author} {\bibinfo {author} {\bibfnamefont {W.}~\bibnamefont
  {Happer}},\ }\href@noop {} {\bibfield  {journal} {\bibinfo  {journal}
  {Reviews of Modern Physics}\ }\textbf {\bibinfo {volume} {44}},\ \bibinfo
  {pages} {169} (\bibinfo {year} {1972})}\BibitemShut {NoStop}%
\bibitem [{\citenamefont {Sansonetti}\ \emph {et~al.}(2013)\citenamefont
  {Sansonetti}, \citenamefont {Martin},\ and\ \citenamefont
  {Young}}]{NISTAtomicBasic}%
  \BibitemOpen
  \bibfield  {author} {\bibinfo {author} {\bibfnamefont {J.~E.}\ \bibnamefont
  {Sansonetti}}, \bibinfo {author} {\bibfnamefont {W.~C.}\ \bibnamefont
  {Martin}}, \ and\ \bibinfo {author} {\bibfnamefont {S.~L.}\ \bibnamefont
  {Young}},\ }\href {\doibase 10.18434/T4FW23} {\emph {\bibinfo {title}
  {Handbook of Basic Atomic Spectroscopic Data (version 1.1.3)}}}\ (\bibinfo
  {publisher} {NIST},\ \bibinfo {year} {2013})\ \bibinfo {note}
  {http://physics.nist.gov/PhysRefData/Handbook/}\BibitemShut {NoStop}%
\bibitem [{\citenamefont {Arimondo}\ \emph {et~al.}(1977)\citenamefont
  {Arimondo}, \citenamefont {Inguscio},\ and\ \citenamefont
  {Violino}}]{arimondo1977experimental}%
  \BibitemOpen
  \bibfield  {author} {\bibinfo {author} {\bibfnamefont {E.}~\bibnamefont
  {Arimondo}}, \bibinfo {author} {\bibfnamefont {M.}~\bibnamefont {Inguscio}},
  \ and\ \bibinfo {author} {\bibfnamefont {P.}~\bibnamefont {Violino}},\
  }\href@noop {} {\bibfield  {journal} {\bibinfo  {journal} {Reviews of Modern
  Physics}\ }\textbf {\bibinfo {volume} {49}},\ \bibinfo {pages} {31} (\bibinfo
  {year} {1977})}\BibitemShut {NoStop}%
\bibitem [{\citenamefont {Weltner}(1989)}]{weltner1989magnetic}%
  \BibitemOpen
  \bibfield  {author} {\bibinfo {author} {\bibfnamefont {W.}~\bibnamefont
  {Weltner}},\ }\href@noop {} {\emph {\bibinfo {title} {Magnetic atoms and
  molecules}}}\ (\bibinfo  {publisher} {Courier Corporation},\ \bibinfo {year}
  {1989})\BibitemShut {NoStop}%
\bibitem [{\citenamefont {Eichler}\ \emph {et~al.}(2002)\citenamefont
  {Eichler}, \citenamefont {M\"uller-Siebert}, \citenamefont {Nettels},
  \citenamefont {Kanorsky},\ and\ \citenamefont
  {Weis}}]{PhysRevLett.88.123002}%
  \BibitemOpen
  \bibfield  {author} {\bibinfo {author} {\bibfnamefont {T.}~\bibnamefont
  {Eichler}}, \bibinfo {author} {\bibfnamefont {R.}~\bibnamefont
  {M\"uller-Siebert}}, \bibinfo {author} {\bibfnamefont {D.}~\bibnamefont
  {Nettels}}, \bibinfo {author} {\bibfnamefont {S.}~\bibnamefont {Kanorsky}}, \
  and\ \bibinfo {author} {\bibfnamefont {A.}~\bibnamefont {Weis}},\ }\href
  {\doibase 10.1103/PhysRevLett.88.123002} {\bibfield  {journal} {\bibinfo
  {journal} {Phys. Rev. Lett.}\ }\textbf {\bibinfo {volume} {88}},\ \bibinfo
  {pages} {123002} (\bibinfo {year} {2002})}\BibitemShut {NoStop}%
\bibitem [{\citenamefont {Moroshkin}\ \emph {et~al.}(2008)\citenamefont
  {Moroshkin}, \citenamefont {Hofer},\ and\ \citenamefont
  {Weis}}]{moroshkin2008atomic}%
  \BibitemOpen
  \bibfield  {author} {\bibinfo {author} {\bibfnamefont {P.}~\bibnamefont
  {Moroshkin}}, \bibinfo {author} {\bibfnamefont {A.}~\bibnamefont {Hofer}}, \
  and\ \bibinfo {author} {\bibfnamefont {A.}~\bibnamefont {Weis}},\ }\href@noop
  {} {\bibfield  {journal} {\bibinfo  {journal} {Physics reports}\ }\textbf
  {\bibinfo {volume} {469}},\ \bibinfo {pages} {1} (\bibinfo {year}
  {2008})}\BibitemShut {NoStop}%
\bibitem [{\citenamefont {Singh}\ \emph {et~al.}(2021)\citenamefont {Singh},
  \citenamefont {Dyre},\ and\ \citenamefont {Pedersen}}]{singh2021solid}%
  \BibitemOpen
  \bibfield  {author} {\bibinfo {author} {\bibfnamefont {A.~N.}\ \bibnamefont
  {Singh}}, \bibinfo {author} {\bibfnamefont {J.~C.}\ \bibnamefont {Dyre}}, \
  and\ \bibinfo {author} {\bibfnamefont {U.~R.}\ \bibnamefont {Pedersen}},\
  }\href@noop {} {\bibfield  {journal} {\bibinfo  {journal} {The Journal of
  Chemical Physics}\ }\textbf {\bibinfo {volume} {154}},\ \bibinfo {pages}
  {134501} (\bibinfo {year} {2021})}\BibitemShut {NoStop}%
\bibitem [{\citenamefont {Goldsborough}\ and\ \citenamefont
  {Koehler}(1964)}]{goldsborough1964electron}%
  \BibitemOpen
  \bibfield  {author} {\bibinfo {author} {\bibfnamefont {J.}~\bibnamefont
  {Goldsborough}}\ and\ \bibinfo {author} {\bibfnamefont {T.}~\bibnamefont
  {Koehler}},\ }\href@noop {} {\bibfield  {journal} {\bibinfo  {journal}
  {Physical Review}\ }\textbf {\bibinfo {volume} {133}},\ \bibinfo {pages}
  {A135} (\bibinfo {year} {1964})}\BibitemShut {NoStop}%
\bibitem [{\citenamefont {Schrimpf}\ \emph {et~al.}(1992)\citenamefont
  {Schrimpf}, \citenamefont {Rosendahl}, \citenamefont {Bornemann},
  \citenamefont {St{\"o}ckmann}, \citenamefont {Faller},\ and\ \citenamefont
  {Manceron}}]{schrimpf1992thermally}%
  \BibitemOpen
  \bibfield  {author} {\bibinfo {author} {\bibfnamefont {A.}~\bibnamefont
  {Schrimpf}}, \bibinfo {author} {\bibfnamefont {R.}~\bibnamefont {Rosendahl}},
  \bibinfo {author} {\bibfnamefont {T.}~\bibnamefont {Bornemann}}, \bibinfo
  {author} {\bibfnamefont {H.-J.}\ \bibnamefont {St{\"o}ckmann}}, \bibinfo
  {author} {\bibfnamefont {F.}~\bibnamefont {Faller}}, \ and\ \bibinfo {author}
  {\bibfnamefont {L.}~\bibnamefont {Manceron}},\ }\href@noop {} {\bibfield
  {journal} {\bibinfo  {journal} {The Journal of chemical physics}\ }\textbf
  {\bibinfo {volume} {96}},\ \bibinfo {pages} {7992} (\bibinfo {year}
  {1992})}\BibitemShut {NoStop}%
\bibitem [{\citenamefont {Vaskonen}\ \emph {et~al.}(1999)\citenamefont
  {Vaskonen}, \citenamefont {Eloranta},\ and\ \citenamefont
  {Kunttu}}]{vaskonen1999trapping}%
  \BibitemOpen
  \bibfield  {author} {\bibinfo {author} {\bibfnamefont {K.}~\bibnamefont
  {Vaskonen}}, \bibinfo {author} {\bibfnamefont {J.}~\bibnamefont {Eloranta}},
  \ and\ \bibinfo {author} {\bibfnamefont {H.}~\bibnamefont {Kunttu}},\
  }\href@noop {} {\bibfield  {journal} {\bibinfo  {journal} {Chemical physics
  letters}\ }\textbf {\bibinfo {volume} {310}},\ \bibinfo {pages} {245}
  (\bibinfo {year} {1999})}\BibitemShut {NoStop}%
\bibitem [{\citenamefont {Carr}\ and\ \citenamefont
  {Purcell}(1954)}]{PhysRev.94.630}%
  \BibitemOpen
  \bibfield  {author} {\bibinfo {author} {\bibfnamefont {H.~Y.}\ \bibnamefont
  {Carr}}\ and\ \bibinfo {author} {\bibfnamefont {E.~M.}\ \bibnamefont
  {Purcell}},\ }\href {\doibase 10.1103/PhysRev.94.630} {\bibfield  {journal}
  {\bibinfo  {journal} {Phys. Rev.}\ }\textbf {\bibinfo {volume} {94}},\
  \bibinfo {pages} {630} (\bibinfo {year} {1954})}\BibitemShut {NoStop}%
\bibitem [{\citenamefont {Slichter}(2013)}]{slichter2013principles}%
  \BibitemOpen
  \bibfield  {author} {\bibinfo {author} {\bibfnamefont {C.~P.}\ \bibnamefont
  {Slichter}},\ }\href@noop {} {\emph {\bibinfo {title} {Principles of magnetic
  resonance}}},\ Vol.~\bibinfo {volume} {1}\ (\bibinfo  {publisher} {Springer
  Science \& Business Media},\ \bibinfo {year} {2013})\BibitemShut {NoStop}%
\bibitem [{\citenamefont {Farfurnik}\ \emph {et~al.}(2018)\citenamefont
  {Farfurnik}, \citenamefont {Jarmola}, \citenamefont {Budker},\ and\
  \citenamefont {Bar-Gill}}]{Farfurnik_2018}%
  \BibitemOpen
  \bibfield  {author} {\bibinfo {author} {\bibfnamefont {D.}~\bibnamefont
  {Farfurnik}}, \bibinfo {author} {\bibfnamefont {A.}~\bibnamefont {Jarmola}},
  \bibinfo {author} {\bibfnamefont {D.}~\bibnamefont {Budker}}, \ and\ \bibinfo
  {author} {\bibfnamefont {N.}~\bibnamefont {Bar-Gill}},\ }\href {\doibase
  10.1088/2040-8986/aaa1bf} {\bibfield  {journal} {\bibinfo  {journal} {Journal
  of Optics}\ }\textbf {\bibinfo {volume} {20}},\ \bibinfo {pages} {024008}
  (\bibinfo {year} {2018})}\BibitemShut {NoStop}%
\bibitem [{\citenamefont {Childress}\ \emph {et~al.}(2006)\citenamefont
  {Childress}, \citenamefont {Dutt}, \citenamefont {Taylor}, \citenamefont
  {Zibrov}, \citenamefont {Jelezko}, \citenamefont {Wrachtrup}, \citenamefont
  {Hemmer},\ and\ \citenamefont {Lukin}}]{childress2006coherent}%
  \BibitemOpen
  \bibfield  {author} {\bibinfo {author} {\bibfnamefont {L.}~\bibnamefont
  {Childress}}, \bibinfo {author} {\bibfnamefont {M.~G.}\ \bibnamefont {Dutt}},
  \bibinfo {author} {\bibfnamefont {J.}~\bibnamefont {Taylor}}, \bibinfo
  {author} {\bibfnamefont {A.}~\bibnamefont {Zibrov}}, \bibinfo {author}
  {\bibfnamefont {F.}~\bibnamefont {Jelezko}}, \bibinfo {author} {\bibfnamefont
  {J.}~\bibnamefont {Wrachtrup}}, \bibinfo {author} {\bibfnamefont
  {P.}~\bibnamefont {Hemmer}}, \ and\ \bibinfo {author} {\bibfnamefont
  {M.}~\bibnamefont {Lukin}},\ }\href@noop {} {\bibfield  {journal} {\bibinfo
  {journal} {Science}\ }\textbf {\bibinfo {volume} {314}},\ \bibinfo {pages}
  {281} (\bibinfo {year} {2006})}\BibitemShut {NoStop}%
\bibitem [{\citenamefont {Abe}\ \emph {et~al.}(2010)\citenamefont {Abe},
  \citenamefont {Tyryshkin}, \citenamefont {Tojo}, \citenamefont {Morton},
  \citenamefont {Witzel}, \citenamefont {Fujimoto}, \citenamefont {Ager},
  \citenamefont {Haller}, \citenamefont {Isoya}, \citenamefont {Lyon},
  \citenamefont {Thewalt},\ and\ \citenamefont {Itoh}}]{PhysRevB.82.121201}%
  \BibitemOpen
  \bibfield  {author} {\bibinfo {author} {\bibfnamefont {E.}~\bibnamefont
  {Abe}}, \bibinfo {author} {\bibfnamefont {A.~M.}\ \bibnamefont {Tyryshkin}},
  \bibinfo {author} {\bibfnamefont {S.}~\bibnamefont {Tojo}}, \bibinfo {author}
  {\bibfnamefont {J.~J.~L.}\ \bibnamefont {Morton}}, \bibinfo {author}
  {\bibfnamefont {W.~M.}\ \bibnamefont {Witzel}}, \bibinfo {author}
  {\bibfnamefont {A.}~\bibnamefont {Fujimoto}}, \bibinfo {author}
  {\bibfnamefont {J.~W.}\ \bibnamefont {Ager}}, \bibinfo {author}
  {\bibfnamefont {E.~E.}\ \bibnamefont {Haller}}, \bibinfo {author}
  {\bibfnamefont {J.}~\bibnamefont {Isoya}}, \bibinfo {author} {\bibfnamefont
  {S.~A.}\ \bibnamefont {Lyon}}, \bibinfo {author} {\bibfnamefont {M.~L.~W.}\
  \bibnamefont {Thewalt}}, \ and\ \bibinfo {author} {\bibfnamefont {K.~M.}\
  \bibnamefont {Itoh}},\ }\href {\doibase 10.1103/PhysRevB.82.121201}
  {\bibfield  {journal} {\bibinfo  {journal} {Phys. Rev. B}\ }\textbf {\bibinfo
  {volume} {82}},\ \bibinfo {pages} {121201} (\bibinfo {year}
  {2010})}\BibitemShut {NoStop}%
\bibitem [{\citenamefont {Yang}\ \emph {et~al.}(2014)\citenamefont {Yang},
  \citenamefont {Burk}, \citenamefont {Widmann}, \citenamefont {Lee},
  \citenamefont {Wrachtrup},\ and\ \citenamefont {Zhao}}]{PhysRevB.90.241203}%
  \BibitemOpen
  \bibfield  {author} {\bibinfo {author} {\bibfnamefont {L.-P.}\ \bibnamefont
  {Yang}}, \bibinfo {author} {\bibfnamefont {C.}~\bibnamefont {Burk}}, \bibinfo
  {author} {\bibfnamefont {M.}~\bibnamefont {Widmann}}, \bibinfo {author}
  {\bibfnamefont {S.-Y.}\ \bibnamefont {Lee}}, \bibinfo {author} {\bibfnamefont
  {J.}~\bibnamefont {Wrachtrup}}, \ and\ \bibinfo {author} {\bibfnamefont
  {N.}~\bibnamefont {Zhao}},\ }\href {\doibase 10.1103/PhysRevB.90.241203}
  {\bibfield  {journal} {\bibinfo  {journal} {Phys. Rev. B}\ }\textbf {\bibinfo
  {volume} {90}},\ \bibinfo {pages} {241203} (\bibinfo {year}
  {2014})}\BibitemShut {NoStop}%
\bibitem [{\citenamefont {Harris}\ \emph {et~al.}(2001)\citenamefont {Harris},
  \citenamefont {Becker}, \citenamefont {De~Menezes}, \citenamefont
  {Goodfellow},\ and\ \citenamefont {Granger}}]{harris2001nmr}%
  \BibitemOpen
  \bibfield  {author} {\bibinfo {author} {\bibfnamefont {R.~K.}\ \bibnamefont
  {Harris}}, \bibinfo {author} {\bibfnamefont {E.~D.}\ \bibnamefont {Becker}},
  \bibinfo {author} {\bibfnamefont {S.~M.~C.}\ \bibnamefont {De~Menezes}},
  \bibinfo {author} {\bibfnamefont {R.}~\bibnamefont {Goodfellow}}, \ and\
  \bibinfo {author} {\bibfnamefont {P.}~\bibnamefont {Granger}},\ }\href@noop
  {} {\bibfield  {journal} {\bibinfo  {journal} {Pure and applied chemistry}\
  }\textbf {\bibinfo {volume} {73}},\ \bibinfo {pages} {1795} (\bibinfo {year}
  {2001})}\BibitemShut {NoStop}%
\bibitem [{\citenamefont {Henry}\ and\ \citenamefont
  {Norberg}(1972)}]{henry1972pulsed}%
  \BibitemOpen
  \bibfield  {author} {\bibinfo {author} {\bibfnamefont {R.}~\bibnamefont
  {Henry}}\ and\ \bibinfo {author} {\bibfnamefont {R.}~\bibnamefont
  {Norberg}},\ }\href@noop {} {\bibfield  {journal} {\bibinfo  {journal}
  {Physical Review B}\ }\textbf {\bibinfo {volume} {6}},\ \bibinfo {pages}
  {1645} (\bibinfo {year} {1972})}\BibitemShut {NoStop}%
\bibitem [{\citenamefont {Sirovich}\ and\ \citenamefont
  {Norberg}(1977)}]{sirovich1977studies}%
  \BibitemOpen
  \bibfield  {author} {\bibinfo {author} {\bibfnamefont {B.}~\bibnamefont
  {Sirovich}}\ and\ \bibinfo {author} {\bibfnamefont {R.}~\bibnamefont
  {Norberg}},\ }\href@noop {} {\bibfield  {journal} {\bibinfo  {journal}
  {Physical Review B}\ }\textbf {\bibinfo {volume} {15}},\ \bibinfo {pages}
  {5107} (\bibinfo {year} {1977})}\BibitemShut {NoStop}%
\bibitem [{\citenamefont {Bar-Gill}\ \emph {et~al.}(2013)\citenamefont
  {Bar-Gill}, \citenamefont {Pham}, \citenamefont {Jarmola}, \citenamefont
  {Budker},\ and\ \citenamefont {Walsworth}}]{NV2013}%
  \BibitemOpen
  \bibfield  {author} {\bibinfo {author} {\bibfnamefont {N.}~\bibnamefont
  {Bar-Gill}}, \bibinfo {author} {\bibfnamefont {L.~M.}\ \bibnamefont {Pham}},
  \bibinfo {author} {\bibfnamefont {A.}~\bibnamefont {Jarmola}}, \bibinfo
  {author} {\bibfnamefont {D.}~\bibnamefont {Budker}}, \ and\ \bibinfo {author}
  {\bibfnamefont {R.~L.}\ \bibnamefont {Walsworth}},\ }\href {\doibase
  10.1038/ncomms2771} {\bibfield  {journal} {\bibinfo  {journal} {Nature
  communications}\ }\textbf {\bibinfo {volume} {4}},\ \bibinfo {pages} {1743}
  (\bibinfo {year} {2013})}\BibitemShut {NoStop}%
\bibitem [{\citenamefont {Abobeih}\ \emph {et~al.}(2018)\citenamefont
  {Abobeih}, \citenamefont {Cramer}, \citenamefont {Bakker}, \citenamefont
  {Kalb}, \citenamefont {Markham}, \citenamefont {Twitchen},\ and\
  \citenamefont {Taminiau}}]{abobeih2018one}%
  \BibitemOpen
  \bibfield  {author} {\bibinfo {author} {\bibfnamefont {M.~H.}\ \bibnamefont
  {Abobeih}}, \bibinfo {author} {\bibfnamefont {J.}~\bibnamefont {Cramer}},
  \bibinfo {author} {\bibfnamefont {M.~A.}\ \bibnamefont {Bakker}}, \bibinfo
  {author} {\bibfnamefont {N.}~\bibnamefont {Kalb}}, \bibinfo {author}
  {\bibfnamefont {M.}~\bibnamefont {Markham}}, \bibinfo {author} {\bibfnamefont
  {D.~J.}\ \bibnamefont {Twitchen}}, \ and\ \bibinfo {author} {\bibfnamefont
  {T.~H.}\ \bibnamefont {Taminiau}},\ }\href@noop {} {\bibfield  {journal}
  {\bibinfo  {journal} {Nature communications}\ }\textbf {\bibinfo {volume}
  {9}},\ \bibinfo {pages} {1} (\bibinfo {year} {2018})}\BibitemShut {NoStop}%
\bibitem [{\citenamefont {Moerner}\ and\ \citenamefont
  {Basche}(1993)}]{moerner1993optical}%
  \BibitemOpen
  \bibfield  {author} {\bibinfo {author} {\bibfnamefont {W.}~\bibnamefont
  {Moerner}}\ and\ \bibinfo {author} {\bibfnamefont {T.}~\bibnamefont
  {Basche}},\ }\href@noop {} {\bibfield  {journal} {\bibinfo  {journal}
  {Angewandte Chemie International Edition in English}\ }\textbf {\bibinfo
  {volume} {32}},\ \bibinfo {pages} {457} (\bibinfo {year} {1993})}\BibitemShut
  {NoStop}%
\bibitem [{\citenamefont {Moerner}(1994)}]{moerner1994examining}%
  \BibitemOpen
  \bibfield  {author} {\bibinfo {author} {\bibfnamefont {W.}~\bibnamefont
  {Moerner}},\ }\href@noop {} {\bibfield  {journal} {\bibinfo  {journal}
  {Science}\ }\textbf {\bibinfo {volume} {265}},\ \bibinfo {pages} {46}
  (\bibinfo {year} {1994})}\BibitemShut {NoStop}%
\bibitem [{\citenamefont {Chambers}\ \emph {et~al.}(2019)\citenamefont
  {Chambers}, \citenamefont {Walton}, \citenamefont {Fairbank}, \citenamefont
  {Craycraft}, \citenamefont {Yahne}, \citenamefont {Todd}, \citenamefont
  {Iverson}, \citenamefont {Fairbank}, \citenamefont {Alamre}, \citenamefont
  {Albert} \emph {et~al.}}]{chambers2019imaging}%
  \BibitemOpen
  \bibfield  {author} {\bibinfo {author} {\bibfnamefont {C.}~\bibnamefont
  {Chambers}}, \bibinfo {author} {\bibfnamefont {T.}~\bibnamefont {Walton}},
  \bibinfo {author} {\bibfnamefont {D.}~\bibnamefont {Fairbank}}, \bibinfo
  {author} {\bibfnamefont {A.}~\bibnamefont {Craycraft}}, \bibinfo {author}
  {\bibfnamefont {D.}~\bibnamefont {Yahne}}, \bibinfo {author} {\bibfnamefont
  {J.}~\bibnamefont {Todd}}, \bibinfo {author} {\bibfnamefont {A.}~\bibnamefont
  {Iverson}}, \bibinfo {author} {\bibfnamefont {W.}~\bibnamefont {Fairbank}},
  \bibinfo {author} {\bibfnamefont {A.}~\bibnamefont {Alamre}}, \bibinfo
  {author} {\bibfnamefont {J.}~\bibnamefont {Albert}},  \emph {et~al.},\
  }\href@noop {} {\bibfield  {journal} {\bibinfo  {journal} {Nature}\ }\textbf
  {\bibinfo {volume} {569}} (\bibinfo {year} {2019})}\BibitemShut {NoStop}%
\end{thebibliography}%

\end{document}